\documentclass[12pt,a4paper]{article}
\pdfoutput=1

\usepackage{jheppub}
\allowdisplaybreaks[1]
\usepackage{amstext,amsfonts,amsthm}

\usepackage{afterpage}
\usepackage{rotating}

\numberwithin{equation}{section}

\usepackage[table]{xcolor}

\usepackage{verbatim}
\usepackage{bbold}
\usepackage{array}
\usepackage{multirow}
\usepackage{booktabs}
\usepackage{longtable}

\definecolor{first}{cmyk}{0,0,0,1}
\definecolor{rest}{gray}{0.6}
\definecolor{second}{rgb}{0,0,1}

\newcommand{\st}{\scriptstyle}

\newcommand{\one}{\ensuremath{\mathbf{1}}}

\newcommand{\bC}{\mathbb{C}}
\newcommand{\bE}{\mathbb{E}}
\newcommand{\bJ}{\mathbb{J}}

\newcommand{\bZ}{\mathbb{Z}}

\newcommand{\cE}{\mathcal{E}}

\newcommand{\cJ}{\mathcal{J}}

\newcommand{\cU}{\mathcal{U}}

\newcommand{\xtoy}{\begin{array}{rcl}
\scriptstyle x_{1}+x_{2}&\scriptstyle\mapsto&\scriptstyle y_{1}\\[-2mm]
\scriptstyle x_{1}x_{2}& \scriptstyle\mapsto& \scriptstyle y_{2}
\end{array}}
\newcommand{\ytox}{\begin{array}{rcl}
\scriptstyle y_{1}&\scriptstyle\mapsto&\scriptstyle x_{1}+x_{2}\\[-2mm]
\scriptstyle y_{2}& \scriptstyle\mapsto& \scriptstyle x_{1}x_{2}
\end{array}}

\newcounter{NumberOfRepeatedColumns}

\newenvironment{SepA}[1]{
    \arrayrulecolor{blue}\left(\begin{array}{#1}
}{%
    \end{array}\right)\arrayrulecolor{black}%
}

\newcommand{\plus}{\pi}
\newcommand{\minus}{\mu}

\hypersetup{
	colorlinks=true, 
	breaklinks=true, 
	linkcolor=green!40!black, 
	citecolor=green!40!black,
	menucolor=green!40!black, 
	urlcolor=green!40!black
	}

\title{Matrix factorisations for rational boundary conditions by defect fusion}

\author[a]{Nicolas Behr}
\author[b]{and Stefan Fredenhagen}
 
 \affiliation[a]{Department of Mathematics,\\
Heriot-Watt University,\\
Riccarton, Edinburgh, EH14 4AS, UK \\
and\\
Maxwell Institute for Mathematical Sciences\\
Edinburgh, UK}
\affiliation[b]{Max-Planck-Institut f{\"u}r Gravitationsphysik,
Albert-Einstein-Institut\\
D-14424 Golm, Germany}

\emailAdd{N.Behr@hw.ac.uk}
\emailAdd{Stefan.Fredenhagen@aei.mpg.de}

\abstract{A large class of two-dimensional $\mathcal{N}=(2,2)$ superconformal
field theories can be understood as IR fixed-points of Landau-Ginzburg
models. In particular, there are rational conformal field theories
that also have a Landau-Ginzburg description. To understand better the
relation between the structures in the rational conformal field theory
and in the Landau-Ginzburg theory, we investigate how rational B-type
boundary conditions are realised as matrix factorisations in the
$SU(3)/U(2)$ Grassmannian Kazama-Suzuki model. As a tool to generate
the matrix factorisations we make use of a particular interface
between the Kazama-Suzuki model and products of minimal models, whose
fusion can be realised as a simple functor on ring modules. This
allows us to formulate a proposal for all matrix factorisations
corresponding to rational boundary conditions in the $SU(3)/U(2)$ model.}

\preprint{AEI-2014-031} 

\begin{document}
\maketitle

\newpage
\section{Introduction}

$\mathcal{N}=(2,2)$ superconformal field theories play an important role as
world-sheet descriptions of superstrings. There are various
constructions and approaches known: the geometric construction as
non-linear sigma model, rational coset constructions (Kazama-Suzuki
models), and the realisation as infrared fixed-point of a
supersymmetric Landau-Ginzburg model (see e.g.\ \cite{Greene:1996cy}
for a review). Each approach has
advantages and disadvantages, in the sense that there are certain
quantities that are easy to compute, and others that are
difficult. For example, in the rational construction one has good
control over the correlation functions, and many quantities can be
determined exactly, but on the other hand, it is hard to compute
deformations of the theory, because the large rational symmetry is
then broken. In contrast to that, in Landau-Ginzburg models
deformations of the superpotential are easily described, but only few quantities
can be computed exactly, namely those that are protected when one
follows the renormalisation group flow to the infrared. It is
therefore desirable to make contact between the different approaches
to combine the advantages and to learn more about the different
descriptions. The connection between the geometric and the
Landau-Ginzburg description is achieved via gauged linear sigma
models~\cite{Witten:1993yc}, and in this way one has obtained a good understanding of the
moduli space of such theories.

We are interested here in the connection between rational theories and
their Landau-Ginzburg realisation. It is known that there is a large
class of supersymmetric coset models that have a Landau-Ginzburg
description, a subclass of the Kazama-Suzuki models~\cite{Kazama:1989qp,Kazama:1988uz}. Within this class
there are the Grassmannian Kazama-Suzuki models that have a
description as cosets $SU(n+1)_{k}/U(n)$. The superpotentials of the
corresponding Landau-Ginzburg theories have been identified
in~\cite{Lerche:1989uy,Gepner:1991gr}, relying on the
identification of the chiral ring of bulk fields.  

In rational theories, one also has a distinguished family of rational
boundary conditions and defects, and it is therefore natural to study
those and to look for their counterparts on the Landau-Ginzburg
side. This has been studied for (products of) minimal models and
orbifolds thereof
in~\cite{Brunner:2003dc,Kapustin:2003rc,Brunner:2005pq,Brunner:2005fv,Enger:2005jk,Keller:2006tf}. In
these models the rational algebras are (products of) super-Virasoro
algebras, so that the algebraic structures are rather simple. A
non-minimal situation has been explored in~\cite{Behr:2010ug}, where
we identified matrix factorisations for some rational boundary
conditions in the $SU(3)/U(2)$ Kazama-Suzuki model. The strategy there
was to identify first some elementary factorisations, and then build
others with the help of the cone construction as tachyon condensates
of elementary ones. This approach, however, cannot be driven very far,
because the cones in question quickly become very complicated.
\smallskip

In this work we want to continue to study the $SU(3)/U(2)$ model, but
following a different approach. The idea is to generate new boundary
conditions by fusing defects onto known boundary conditions. If we
have identified the appropriate defects as matrix factorisations, we
can use them to generate new matrix factorisations for boundary
conditions from known ones by taking tensor products of matrix factorisations.

To identify matrix factorisations for defects, we make use of an
interface between the $SU(3)/U(2)$ Kazama-Suzuki model and the product of two minimal
models that we introduced in~\cite{Behr:2012xg}. The fusion of this variable
transformation interface to a matrix factorisation has a simple
operator-like description: it can be implemented by a simple operation
acting individually on each entry of the matrix factorisation. 

Fusing this interface to a matrix factorisation in the minimal
models results in a matrix factorisation for the Kazama-Suzuki
model. This interface then allows us to identify a matrix
factorisation for a particular rational topological defect in the
Kazama-Suzuki model. Fusing this defect to the matrix factorisations
identified in~\cite{Behr:2010ug}, we generate matrix factorisations 
for all rational boundary conditions.
\smallskip

The plan of the paper is as follows. In section~\ref{sec:MFs} we
review matrix factorisations for B-type boundary conditions in
Landau-Ginzburg models and the variable transformation interface
between the Kazama-Suzuki model and products of minimal
models. Section~\ref{sec:CFT} gives an introduction to the conformal
field theory description of Kazama-Suzuki models. We discuss rational boundary
conditions and renormalisation group flows between them. Defects and
their fusion to boundary conditions are briefly reviewed. After these
preparations we discuss in section~\ref{sec:rationalfactorisations}
the construction of matrix factorisations for rational boundary
conditions. We show how the factorisations of~\cite{Behr:2010ug} can
be obtained from permutation factorisations in the product of two
minimal models with the help of the variable transformation
interface. We also discuss how the interface relates the computation of RR-charges in
Kazama-Suzuki models to computations in minimal models. Finally the
interface is used to construct a certain topological defect in the
Landau-Ginzburg description that then allows us to algorithmically determine matrix
factorisations for all rational boundary conditions. We compute a
large class of them explicitly, and formulate a concrete proposal for
all such factorisations. We have collected
some of the more technical steps in the appendix.

\section{Matrix factorisations and variable transformation
interfaces}\label{sec:MFs}

In this section we introduce the description of B-type boundary
conditions in Landau-Ginzburg models as matrix factorisations. We then
review the construction of variable transformation interfaces, and
discuss in detail the interface between the $SU(3)/U(2)$ Kazama-Suzuki
model and the product of two minimal models.

\subsection{Matrix factorisations in Landau-Ginzburg models}

B-type boundary conditions in $\mathcal{N}=(2,2)$ supersymmetric
Landau Ginzburg models can be described by matrix factorisations $Q$
of the superpotential $W$
(see~\cite{Kontsevich:unpublished,Kapustin:2002bi,Orlov:2003yp,Brunner:2003dc,Kapustin:2003ga}). We
want to consider a polynomial superpotential $W(x_{1},\dotsc ,x_{n})$,
and the factorisation $Q$ is then a polynomial square matrix of the form
\begin{equation}
Q=\begin{pmatrix}
0 & Q^{(1)}\\
Q^{(0)} & 0 
\end{pmatrix}
\end{equation}
such that
\begin{equation}
Q^{2} = W \cdot \one \ .
\end{equation}
The spectrum of chiral primary boundary fields is encoded in terms of
morphisms between matrix factorisations. Let $Q_{1}$ and $Q_{2}$ be
two matrix factorisations of size $2q_{1}$ and $2q_{2}$,
respectively. $Q_{i}$ implements an endomorphism on $R^{2q_{i}}$, where
$R=\mathbb{C}[x_{1},\dotsc ,x_{n}]$ is the polynomial ring in the
variables $x_{1},\dotsc ,x_{n}$. There is a natural $\mathbb{Z}_{2}$
grading on these free modules, $R^{2q_{i}}=R^{q_{i}}\oplus R^{q_{i}}$,
such that $Q_{i}$ defines an odd map. Also morphisms $\phi_{n}$
between $Q_{1}$ and $Q_{2}$ come with a $\mathbb{Z}_{2}$ degree
$n$. They are given by even ($n=0$) or odd ($n=1$) homomorphisms from $R^{2q_{1}}$ to $R^{2q_{2}}$
that satisfy the closure condition
\begin{equation}
Q_{2}\,\phi_{n} - (-1)^{n} \phi_{n}\, Q_{1} = 0 \ .
\end{equation}
In addition, two morphisms that differ by an exact morphism of the form
\begin{equation}\label{exact}
\tilde{\phi}_{n} = Q_{2}\,\psi + (-1)^{n}\psi \, Q_{1}
\end{equation}
are identified. 

If for two matrix factorisations $Q_{1}$, $Q_{2}$ there is a homomorphism
$\phi_{0}$ between $Q_{1}$ and $Q_{2}$, and a homomorphism
$\psi_{0}$ between $Q_{2}$ and $Q_{1}$, such that
$\phi_{0}\circ \psi_{0}$ and $\psi_{0}\circ \phi_{0}$
coincide with the identity up to exact terms~\eqref{exact}, then we
say that these two matrix factorisations are equivalent. 

In particular if the factorisations $Q_{1}$ and $Q_{2}$ are of the same size and are related by
a similarity transformation $\cU$,
\begin{equation}
Q_{2} = \cU \cdot Q_{1} \cdot \cU^{-1} \ ,
\end{equation}  
then $Q_{1}$ and $Q_{2}$ are equivalent with $\phi_{0}=\cU$ and $\psi_{0}=\cU^{-1}$.

Given two factorisations $Q_{1}$ and $Q_{2}$ and an odd morphism
$\phi_{1}$ from $Q_{1}$ to $Q_{2}$, one can build a new factorisation
$C (Q_{1},Q_{2};\phi_{1})$ by the so-called cone construction that is
related to the process of tachyon condensation (see e.g.\
\cite{Herbst:2004zm,govindarajan:2005im}),
\begin{equation}
C (Q_{1},Q_{2};\phi_{1}) = \begin{pmatrix}
Q_{1} & 0\\
\phi_{1} & Q_{2}
\end{pmatrix} \ .
\end{equation}

\subsection{Variable transformation interfaces}

We can describe B-type interfaces between Landau-Ginzburg models with
superpotentials $W^{x}(x_{1},\dotsc ,x_{m})$ and $W^{y} (y_{1},\dotsc
,y_{n})$ by matrix factorisations of the difference $W^{x}-W^{y}$ of
the superpotentials~\cite{Brunner:2007qu} (see
also~\cite{Kapustin:2004df,Khovanov:2004}). They can be fused to other
matrix factorisations by means of the tensor product of matrix
factorisations~\cite{Yoshino:1998,Khovanov:2004}.

If the two superpotentials are related to each other by a variable transformation,
\begin{equation}\label{vartrans}
y_{j} \mapsto Y_{j} (x_{1},\dotsc ,x_{m}) \ ,
\end{equation}
that expresses the $y_{j}$ as polynomials in the variables $x_{i}$, such that 
\begin{equation}
W^{x} (x_{1},\dotsc ,x_{m}) = W^{y}(Y_{1} (x_{1},\dotsc
,x_{m}),\dotsc ,Y_{n} (x_{1},\dotsc ,x_{m})) \ ,
\end{equation}
there is a particular \emph{variable transformation interface}
$_{y}I_{x}$ that we introduced in~\cite{Behr:2012xg}. The fusion of
this interface to other matrix factorisations can be described in a
simple way as we will review in the following.

If we denote the polynomial rings in $x_{i}$ and $y_{i}$ variables by
$S$ and $R$, respectively, the variable
transformation~\eqref{vartrans} defines a ring homomorphism $Y$,
\begin{equation}
Y : R \to S \quad ,\quad Y : p (y_{1},\dotsc ,y_{n}) \mapsto p
(Y_{1} (x_{1},\dotsc ,x_{m}),\dotsc ,Y_{n} (x_{1},\dotsc ,x_{m})) \ .
\end{equation}
Using this homomorphism we can view $S$ as an $(S,R)$-bimodule
$_{S}S_{R}$ or as an $(R,S)$-bimodule $_{R}S_{S}$.  This defines two
functors, the \emph{extension of scalars} $Y^{*}$ maps $R$-modules to
$S$-modules by tensoring with $_{S}S_{R}$, and the \emph{restriction of
scalars} $Y_{*}$ maps $S$- to $R$-modules by tensoring with
$_{R}S_{S}$. 

Let us discuss the first one, $Y^{*}$, more explicitly. First we
observe that this functor maps finite rank free $R$-modules to
finite rank free $S$-modules of the same rank,
\begin{equation}
_{S}S_{R} \otimes_{R} \big({}_{R}R\oplus \dotsb \oplus {}_{R}R \big) \cong
{}_{S}S \oplus \dotsb \oplus {}_{S}S \ .
\end{equation}
A homomorphism between finite rank free $R$-modules, which can be viewed as a
matrix with polynomial entries in the variables $y_{i}$, is mapped to
the homomorphism between $S$-modules that is obtained by replacing all
variables $y_{i}$ by the polynomials $Y_{i} (x_{1},\dotsc ,x_{m})$. So
it acts by replacement of variables: it takes polynomial matrices in variables $y_{j}$ and
maps them to polynomial matrices in variables $x_{i}$.

The second one, $Y_{*}$, maps an $S$-module to an $R$-module by
tensoring it with $_{R}S_{S}$,
\begin{equation}
_{S}M \mapsto {}_{R}S_{S}\otimes_{S}{}_{S}M \ .
\end{equation}
This is in general not a finite rank free $R$-module, even if 
$_{S}M$ was a finite rank free $S$-module. If on the other hand
$_{R}S$ as an $R$-module is free and of finite rank,
\begin{equation}
\rho : {}_{R}R^{\oplus r}  \xrightarrow{\sim}   {}_{R}S \ ,
\end{equation}
with $\rho$ an $R$-module isomorphism, then a free $S$-module $_{S}M$
of rank $d$ is mapped to a free $R$-module of rank $r\cdot d$. In this
case, its action on homomorphisms can also be described very
concretely: given any homomorphism $\phi$ of free $S$-modules of
finite rank, we can represent it by a matrix whose entries $\phi_{ij}$
are polynomials in $S$. The homomorphism between the images of the
modules under $Y_{*}$ is then described by the matrix that is
obtained by replacing each entry $\phi_{ij}$ by a $r\times r$-block
that describes the map $\rho^{-1}\circ \phi_{ij}\circ \rho$. Therefore
the functor $Y_{*}$ maps matrices in the variables $x_{i}$ to (in
general larger) matrices in the variables $y_{j}$.

To summarise, we have introduced two functors that on polynomial entries act as
\begin{align}
Y^{*} (p (y_{1},\dotsc ,y_{n})) &= p \big(Y_{1} (x_{1},\dotsc
,x_{m}),\dotsc ,Y_{n} (x_{1},\dotsc ,x_{m}) \big)\\
Y_{*} ( p (x_{1},\dotsc ,x_{m})) & = \rho^{-1}\circ p\circ \rho \
.
\label{actionofyUx}
\end{align}
These two functors describe the fusion of the variable transformation
interface $_{y}I_{x}$: fusing it to the left, it acts by replacement
of variables (i.e.\ via $Y^{*}$), fusing it to the right it acts by $Y_{*}$.
\smallskip

The simplest example of a variable transformation interface is
obtained if the rings are the same, $S=R$, and the map $Y=\sigma$ is an
automorphism of $R$. In this case, $Y^{*}$ acts by replacing variables
according to $Y$, whereas the action of $Y_{*}$ is given by the
inverse $Y^{-1}$. In case the two superpotentials are the same, and
$\sigma$ is a symmetry of $W$, these interfaces are also known as
group-like defects or symmetry defects~\cite{Frohlich:2006ch,Brunner:2007qu,Carqueville:2009ev}.

\subsection{Kazama-Suzuki models}\label{sec:ks-minmod-functor}

We now come to our key example, which will be important for the rest
of this paper. These are the \emph{Grassmannian Kazama-Suzuki models}
$SU(n+1)/U(n)$, where we will be interested in particular in the case
$n=2$. 

\noindent For general $n\geq 1$, we consider the superpotential
\begin{equation}
W^{x}_{n;k} (x_{1},\dotsc ,x_{n}) = x_{1}^{k+n+1}+\dotsb +x_{n}^{k+n+1}\ ,
\end{equation}
where $n,k\geq 1$ are integers. As $W^{x}$ is completely symmetric in
$x_{1},\dotsc,x_{n}$, we can express it in terms of the elementary
symmetric polynomials
\begin{equation}
Y_{j} (x_{1},\dotsc ,x_{n}) = \sum_{1\leq i_{1}<\dotsb <i_{j}\leq n} x_{i_{1}}\cdot \dotsb
\cdot x_{i_{j}} \quad ,\ j=1,\dotsc ,n \ ,
\end{equation} 
to obtain a superpotential $W^{y}$ in variables $y_{1},\dotsc ,y_{n}$ such that
\begin{equation}
W^{y}_{n;k} (Y_{1} (x_{1},\dotsc ,x_{n}),\dotsc ,Y_{n} (x_{1},\dotsc
,x_{n})) = W^{x}_{n;k} (x_{1},\dotsc ,x_{n}) \ .
\end{equation}
The superpotential $W^{x}$ describes the tensor product of $n$ minimal
models, whereas $W^{y}$ describes the $SU(n+1)/U(n)$ Kazama-Suzuki
model (see~\cite{Gepner:1988wi,Lerche:1989uy}). We are now
precisely in the setup of the previous subsection, and we can define a
variable transformation interface $_{x}I_{y}$ between these models. It
acts on the right just by replacing the variables $y_{j}$ by
$Y_{j}(x_{1},\dotsc ,x_{n})$. To understand its behaviour on the left,
i.e.\ its action on the $x$-variables, we have to understand the
structure of $S=\mathbb{C}[x_{1},\dotsc,x_{n}]$ as a module over
$R=\mathbb{C}[y_{1},\dotsc ,y_{n}]$. In the following we want to
restrict to the case $n=2$. We choose the explicit
$R$-module isomorphism $\rho$ between $R\oplus R$ and $_{R}S$ as
\begin{equation}
\rho : \big(p_{1} (y_{1},y_{2}),p_{2} (y_{1},y_{2}) \big) \mapsto
p_{1} (x_{1}+x_{2},x_{1}x_{2}) + (x_{1}-x_{2})
p_{2} (x_{1}+x_{2},x_{1}x_{2}) \ .
\end{equation}
The inverse is then given by 
\begin{equation}
\rho^{-1}: p (x_{1},x_{2}) \mapsto \bigg(p_{S}
(x_{1},x_{2})\Big|_{y},\frac{1}{x_{1}-x_{2}}p_{A}
(x_{1},x_{2})\Big|_{y}\bigg)\ ,
\end{equation}
where 
\begin{equation}
p_{S/A} (x_{1},x_{2}) =\frac{1}{2}\big(p(x_{1},x_{2})\pm
p(x_{2},x_{1})\big)\ ,
\end{equation}
and for a symmetric polynomial $q(x_{1},x_{2})$
we denote by $q(x_{1},x_{2})|_{y}$ the polynomial in $y$-variables
from which one obtains $q(x_{1},x_{2})$ when one replaces $y_{i}$ by
$Y_{i}(x_{1},x_{2})$.  

The functor $Y_{*}$ sends an $S$-module of rank $r$ to
an $R$-module of rank $2r$. On homomorphisms it acts by replacing each
polynomial entry by a $2\times 2$ matrix. With the explicit
isomorphism $\rho$ given above, the action of $Y_{*}$ on a
polynomial $p(x_{1},x_{2})$ can be determined
from~\eqref{actionofyUx}, and it is given by
\begin{equation}\label{ksexample-actionofyUx}
Y_{*}:p \mapsto \begin{pmatrix}
p_{S}\big|_y & (x_{1}-x_{2})p_{A} \big|_{y}\\
\frac{p_{A}}{x_{1}-x_{2}}\big|_{y} & p_{S}\big|_{y}
\end{pmatrix} \ .
\end{equation}
This variable transformation interface can then be used to relate
defects and boundary conditions in Kazama-Suzuki models to those in
minimal models. It lies at the heart of the constructions in this paper.

\section{Boundaries and Defects in Kazama-Suzuki models}\label{sec:CFT}

In this section we review the construction of rational boundary conditions
in Grassmannian Kazama-Suzuki model with emphasis on the
model based on the coset $SU(3)/U(2)$. We also discuss
renormalisation group flows of boundary conditions, and topological
defects and their fusion to boundaries.

\subsection{Bulk theory}

Kazama-Suzuki models~\cite{Kazama:1989qp,Kazama:1988uz} are rational
$\mathcal{N}=(2,2)$ superconformal field theories that are constructed as cosets 
\begin{equation}
\frac{G_{k}\times SO (d)_{1}}{H}\ ,
\end{equation}
where $d$ is the difference between the dimension of the simple Lie
group $G$ and the dimension of its regularly embedded subgroup
$H$. The integer $k$ is the level, and for $\mathcal{N}=(2,2)$
supersymmetry, the geometric space $G/H$ has to be K{\"a}hler.
A particularly interesting class of such models are the Grassmannian
models based on $G=SU(n+1)$ and $H=U(n)$, and in this work we specify
the model further by considering the case $n=2$.

In the following we briefly review the spectrum of the $SU(3)/U(2)$
model. More details can be found e.g.\ in~\cite{Behr:2010ug}. 
The primary fields (w.r.t.\ the bosonic subalgebra of the chiral
symmetry algebra) are labelled by tuples $(\Lambda ,\Sigma;\lambda ,\mu)$ where 
\begin{itemize}
\item $\Lambda=(\Lambda_{1},\Lambda_{2})$ is an $su(3)$ highest weight
($\Lambda_{1},\Lambda_{2}$ being the non-negative integer Dynkin
labels) satisfying $\Lambda_{1}+\Lambda_{2}\leq k$, and it labels a (unitary irreducible)
representation of the affine Lie algebra $su(3)_{k}$,
\item $\Sigma\in\{0,v,s,c\}$ labels representations of $so(4)_{1}$
(with the corresponding representations being the trivial
representation, vector, spinor and conjugate spinor),
\item $\lambda$ is a highest weight of $su(2)$, with $0\leq \lambda\leq k+1$ 
labelling a representation of the affine $su(2)_{k+1}$,
\item $\mu$ is an integer modulo $6(k+3)$ labelling
representations of $u(1)_{6(k+3)}$.
\end{itemize}
There is a selection rule on the allowed labels that reads
\begin{equation}
\frac{\Lambda_{1}+2\Lambda_{2}}{3} + \frac{|\Sigma |}{2}
-\frac{\lambda}{2}+\frac{\mu}{6} \in \mathbb{Z} \ ,
\end{equation}
where $|\Sigma|=0$ for $\Sigma =0,v$ and $|\Sigma|=1$ for
$\Sigma=s,c$. Finally, tuples are identified according to
\begin{equation}
((\Lambda_{1},\Lambda_{2}),\Sigma;\lambda,\mu) \sim
((k-\Lambda_{1}-\Lambda_{2},\Lambda_{1}),v\times
\Sigma;k+1-\lambda,\mu+ (k+3))\ , 
\end{equation}
where $v\times \cdot$ denotes the fusion with the vector
representation, which exchanges on the one hand $0$ and $v$, and on
the other hand $s$ and $c$.

In the spectrum there are chiral primary fields corresponding to the tuples
\begin{equation}\label{chiralprimary}
((\Lambda_{1},\Lambda_{2}),0;\Lambda_{1},\Lambda_{1}+2\Lambda_{2}) \ ,
\end{equation}
and they can be labelled by representations 
$(\Lambda_{1},\Lambda_{2})$ of $su(3)$.

\subsection{Boundary conditions}

According to how the supercurrents are glued at the boundary of the
world-sheet we distinguish between A-type and B-type gluing
conditions~\cite{Ooguri:1996ck}. Here we are only interested in B-type
gluing conditions. Rational boundary conditions can be constructed
following the Cardy construction~\cite{Cardy:1989ir}. In the diagonal
$SU(3)/U(2)$ coset model, maximally symmetric B-type boundary states
$|L,S;\ell \rangle$ are labelled by two integers $L,\ell$ with $0\leq
L\leq \lfloor \frac{k}{2}\rfloor$, $0\leq \ell\leq k+1$, and an
$so(4)_{1}$ representation $S$ (see e.g.\ \cite{Behr:2010ug}, and
also~\cite{Ishikawa:2003kk} for a general discussion of twisted
boundary states in Kazam-Suzuki models). Here, $\lfloor x\rfloor$
denotes the greatest integer smaller or equal $x$.  Choosing a
particular sign in the gluing condition for the supercurrents, we can
restrict to $S=0,v$. We introduce the notation
\begin{equation}
|L,\ell \rangle := |L,0;\ell\rangle \quad \text{and}\quad 
\overline{|L,\ell \rangle} := |L,v;\ell \rangle \ .
\end{equation}
Because of field identifications and selection rules, we have to identify
\begin{equation}\label{bsidentification}
|L,\ell\rangle \equiv \overline{|L,k+1-\ell\rangle} \ .
\end{equation}
The boundary spectrum is given by ($q=e^{2\pi i\tau}$,
$\tilde{q}=e^{-2\pi i/\tau}$)
\begin{multline}
\langle L,\ell|q^{\frac{1}{2} (L_{0}+\bar{L}_{0})-\frac{c}{24}}|L',\ell'\rangle\\
= \sum_{[\Lambda,\Sigma;\lambda,\mu]} n_{\Lambda L}{}^{L'} \big(
N^{(k+1)}_{\lambda\ell}{}^{\ell'}\delta_{\Sigma,0} 
+ N^{(k+1)}_{\lambda(k+1-\ell)}{}^{\ell'}\delta_{\Sigma,v}\big)\chi_{(\Lambda,\Sigma;\lambda,\mu)} (\tilde{q}) \big)\ .
\end{multline}
Here, the sum only goes over equivalence classes
of bulk labels, and $N^{(k+1)}$ denotes the fusion rules of
$su(2)_{k+1}$, $N^{\text{so}}$ the fusion rules of $so(4)_{1}$, and 
\begin{equation}\label{twistedfusionrules}
n_{\Lambda L}{}^{L'} = \sum_{\lambda} b^{\Lambda}_{\lambda}
\left(N^{(k+1)}_{\lambda L}{}^{L'} - N^{(k+1)}_{(k+1-\lambda)L}{}^{L'} \right)
\end{equation}
are twisted fusion rules of $su(3)_{k}$ (see e.g.\ \cite{Gaberdiel:2002qa}). In the last expression the
branching rules $b^{\Lambda}_{\lambda}$ of the decomposition of $su(3)$ representations
$\Lambda =(\Lambda_{1},\Lambda_{2})$ into
representations $\lambda$ of its regularly embedded subalgebra $su(2)$ appear.
We will later need the branching rules that describe how an $su(3)$
representation $(\Lambda_{1},\Lambda_{2})$ decomposes into
representations $(\lambda;\mu)$ of $su(2)\oplus u(1)$,
\begin{equation}\label{branching}
(\Lambda_{1},\Lambda_{2}) \to \sum_{\lambda,\mu}
b^{\Lambda}_{(\lambda;\mu)} (\lambda;\mu) = 
\sum_{\gamma_{1}=0}^{\Lambda_{1}}\sum_{\gamma_{2}=0}^{\Lambda_{2}}
\big(\gamma_{1}+\gamma_{2}; 3(\gamma_{1}-\gamma_{2}) 
+ 2(\Lambda_{2}-\Lambda_{1})\big) \ .
\end{equation}
From this we directly read off the branching needed
in~\eqref{twistedfusionrules} by ignoring the $u(1)$ label $\mu$.

\subsection{Boundary renormalisation group flows}

When relevant boundary fields are present, one can study the boundary
renormalisation group flows induced by those fields. Such boundary
flows have been studied in general cosets in the limit of large
levels~\cite{Fredenhagen:2001nc,Fredenhagen:2001kw,Fredenhagen:thesis}. There
is one class of flows that is conjectured to be present at all
levels~\cite{Fredenhagen:2002qn,Fredenhagen:2003xf,Bachas:2009mc}, which we will
briefly describe here. 

Applied to the $SU(3)/U(2)$ Kazama-Suzuki models, the rule
of~\cite{Fredenhagen:2002qn,Fredenhagen:2003xf} predicts the following
renormalisation group flows:
\begin{equation}
\sum_{\lambda,\ell'} b^{\Lambda^{+}}_{\lambda}\,
N^{(k+1)}_{\lambda\ell}{}^{\ell'}\,|L,\ell'\rangle
\longrightarrow \sum_{L'} n_{\Lambda L}{}^{L'}\, |L',\ell \rangle \ ,
\end{equation}
where $\Lambda=(\Lambda_{1},\Lambda_{2})$ is an arbitrary highest
weight with $\Lambda_{1}+\Lambda_{2}\leq k$ labelling a representation
of $su(3)_{k}$, and $\Lambda^{+}=(\Lambda_{2},\Lambda_{1})$ is the
conjugate representation. $b^{\Lambda}_{\lambda}$ denotes the
branching of the $su(3)$ representation $\Lambda$ into $su(2)$
representations $\lambda$ (see~\eqref{branching}). The field that
induces this flow is a linear combination of fields labelled by
$((0,0),0;1,\pm 3)$.

A simple example of such a flow is given by $\Lambda = (1,0)$, and it reads
\begin{multline}\label{simplestflow}
|L,\ell -1\rangle + |L,\ell\rangle + |L,\ell +1\rangle \\
\longrightarrow \left\{\begin{array}{ll}
|L-1,\ell \rangle + |L,\ell\rangle + |L+1,\ell\rangle &
\text{for}\ 
L\not= \frac{k}{2}\\
|L-1,\ell\rangle & \text{for}\ L=\frac{k}{2} \ .
\end{array} \right.
\end{multline}
If a label happens to lie outside the allowed range, the corresponding
boundary state has to be omitted (e.g.\ for $\ell=0$ the first state
on the left hand side can be left out).

A nice outcome of this flow rule is that one can obtain all boundary
states from a subset of states by perturbing suitable superpositions
of boundary states. Successively using the flow~\eqref{simplestflow}
one can e.g.\ start from the states $|0,\ell\rangle$ and obtain all others.

\subsection{Defects and fusion}\label{sec:RCFTdefects}

We can also study topological defects in these models, and here we
will focus on defects with B-type gluing conditions for the
supercurrents. The rational defects carry the same labels as the bulk fields,
$D_{[\Lambda,\Sigma ;\lambda,\mu]}$ \cite{Petkova:2000ip}. By fixing the sign in the gluing
condition for the supercurrents we can restrict the set of defects to
those with $\Sigma =0,v$.

Topological defects can be fused to
boundaries~\cite{Petkova:2000ip,Graham:2003nc}. Using a B-type defect,
a B-type boundary condition is transformed into a superposition of
B-type boundary conditions,
\begin{equation}\label{eq:FSdefBrane}
D_{[\Lambda ,0;\lambda ,\mu]} |L,\ell\rangle  = 
\sum n_{\Lambda L}{}^{L'}\, N^{(k+1)}_{\lambda \ell}{}^{\ell'} \,
|L',\ell'\rangle \ .
\end{equation}
Defects that only differ in the label $\mu$ have an identical effect
on B-type boundary conditions.

As an example consider the defect $D_{[(0,0),0;1,3]}$. Fusing this
defect to boundary conditions is described by
\begin{equation}\label{examplefusion}
D_{[(0,0),0;1,3]} |L,\ell \rangle = |L,\ell +1\rangle + 
|L,\ell -1\rangle \ ,
\end{equation}
where the last boundary condition is omitted if $\ell =0$. Therefore,
starting from $|L,0\rangle$ one can generate all other boundary
conditions by fusing $D_{[(0,0),0;1,3]}$.

\section{Matrix factorisations for rational boundary
conditions}\label{sec:rationalfactorisations}

In this section we want to discuss matrix factorisations of the
Landau-Ginzburg superpotential $W^{y}_{2;k}$ that leads to the
$SU(3)/U(2)$ Kazama-Suzuki model. In particular we want to identify
those factorisations that correspond to rational boundary conditions in
the conformal field theory. 

We first review the identification of some of the rational boundary
conditions as polynomial factorisations (i.e. where the matrix
factorisations $Q$ are $2\times 2$-matrices) \cite{Behr:2010ug}, and how one can
obtain some higher factorisations via the cone construction. Then we will discuss
how one can employ defects for a systematic construction of all matrix
factorisations corresponding to rational boundary conditions.

\subsection{Polynomial factorisations}\label{sec:polynomialMF}

The superpotential of the $SU(3)/U(2)$ Kazama-Suzuki model is given by
\begin{align}
W^{y}_{2;k} (y_{1},y_{2}) &= \big(
x_1^{k+3}+x_2^{k+3}\big)\Big\vert_{\xtoy}
= \prod_{j=0}^{k+2}\big(x_{1} - \eta^{2j+1}x_{2} \big)\Big\vert_{\xtoy}\nonumber\\
&= \prod_{j=0}^{\lfloor\frac{k+1}{2} \rfloor} (y_{1}^{2}-\delta
_{j}y_{2}) \cdot \left\{\begin{array}{ll}
y_{1} & \text{for}\ k\ \text{even}\\
1    & \text{for}\ k\ \text{odd} \ ,
\end{array} \right.\label{eq:WKSdef}
\end{align}
where 
\begin{equation}
\eta = e^{i\pi /(k+3)} \quad ,\quad 
\delta_{j} = \frac{\big(1+\eta^{2j+1}\big)^{2}}{\eta^{2j+1}} \ .
\end{equation}
The product form of the superpotential allows us to easily write down
factorisations $Q^{(1)}\cdot Q^{(0)}=W^{y}_{2;k}$ with polynomials
$Q^{(1)}$ and $Q^{(0)}$. Among those polynomial factorisations we
could identify in~\cite{Behr:2010ug} those that correspond to rational boundary
conditions. One class that can be identified in this way consists of
the boundary conditions $|L,0\rangle$, and the associated
factorisations are
\begin{equation}\label{eq:QLzeroMFs}
Q_{|L,0\rangle} = \begin{pmatrix}
0 &  \cJ_{|L,0\rangle}\\
\cJ_{\overline{|L,0\rangle}} & 0
\end{pmatrix}\ ,
\end{equation}
with
\begin{equation}\label{cJs}
\cJ_{|L,0\rangle}= \prod_{j=0}^{L} \cJ_{j} \quad ,\quad \cJ_{j} =
y_{1}^{2}-\delta_{j}y_{2} \quad ,\quad \cJ_{\overline{|L,0\rangle}}=
\frac{W^{y}_{2;k}}{\cJ_{|L,0\rangle}} \ .
\end{equation}
The identification in~\cite{Behr:2010ug} is based on the comparison of the
spectra of chiral primary fields, and of the RR-charges.

For even $k$ there is another class of rational boundary conditions
that have a description in terms of polynomial factorisations. These
are the boundary conditions $|\frac{k}{2},\ell \rangle$ -- details can
be found in~\cite{Behr:2010ug}. 
\smallskip

In section~\ref{sec:ks-minmod-functor} we introduced the variable
transformation interface~$_{y}I_{x}$ between the $SU(3)/U(2)$
Kazama-Suzuki model and two copies of minimal models at level
$k+1$. Let us briefly discuss how one can obtain the factorisations
$Q_{|L,0\rangle}$ in the Kazama-Suzuki model from factorisations in
the product of minimal models by interface fusion.  The simplest
factorisations in the product of two minimal models are the polynomial
factorisations, which are called permutation
factorisations~\cite{Brunner:2005fv} (see
also~\cite{Ashok:2004zb,Ashok:2004xq}). A subset of those corresponds
to rational boundary states, namely the permutation boundary
states~$|L,M\rangle_{\text{perm}}$, which are labelled by two numbers,
$L=0,\dotsb,k+1 $ and $M$ being an integer identified modulo $2k+6$,
such that $L+M$ is even. In~\cite{Brunner:2005fv} these have been
identified with the factorisations
\begin{align}
Q_{|L,M\rangle_{\text{perm}}}&=\begin{pmatrix}
0& Q^{(1)}_{|L,M\rangle_{\text{perm}}}\\
Q^{(0)}_{|L,M\rangle_{\text{perm}}} & 0
\end{pmatrix}\nonumber\\
&=\begin{pmatrix}
0 & {\displaystyle\prod_{j=-\frac{M+L}{2}-1}^{-\frac{M-L}{2}-1}} \big(x_{1}-\eta^{2j+1}x_{2}\big)\\
{\displaystyle\prod_{-\frac{M-L}{2}}^{k+1-\frac{M+L}{2}}} \big(x_{1}-\eta^{2j+1}x_{2}\big) & 0
\end{pmatrix}\ .
\label{permutationMF}
\end{align}
Let us now fuse the interface $_{y}I_{x}$ onto the factorisation
$Q_{|2L,0\rangle_{\text{perm}}}$. We first note that we can rewrite the
product that appears in $Q^{(1)}_{|2L,0\rangle_{\text{perm}}}$ as
\begin{equation}
Q^{(1)}_{|2L,0\rangle_{\text{perm}}} =
\big(x_{1}-\eta^{-2L-1}x_{2}\big)\prod_{j=0}^{L-1} \cJ_{j}
(y_{1},y_{2})\Big|_{\ytox} \ .
\end{equation}
The effect of fusing $_{y}I_{x}$ is given by the functor $Y_{*}$ defined
in~\eqref{ksexample-actionofyUx}. When we apply it to
$Q^{(1)}_{|2L,0\rangle_{\text{perm}}}$, we obtain
\begin{align}
Y_{*} \big(Q^{(1)}_{|2L,0\rangle_{\text{perm}}}\big) &=
\prod_{j=0}^{L-1}\cJ_{j} (y_{1},y_{2})\cdot \begin{pmatrix}
\frac{1}{2} \big(1-\eta^{-L-1}\big) y_{1} &
\frac{1}{2} \big(1+\eta^{-L-1}\big) \big(y_{1}^{2}-4y_{2}\big)\\
 \frac{1}{2} \big(1+\eta^{-L-1}\big) & \frac{1}{2} \big(1-\eta^{-L-1}\big) y_{1}
\end{pmatrix} \nonumber\\
&\to \begin{pmatrix}
 \prod_{j=0}^{L}\cJ_{j} (y_{1},y_{2}) & 0 \\
0 & \prod_{j=0}^{L-1}\cJ_{j} (y_{1},y_{2})
\end{pmatrix} \ ,
\end{align}
where we performed a similarity transformation in the second step. We thus see that
\begin{equation}\label{KS-bc-from-minmod}
_{y}I_{x}\otimes Q_{|2L,0\rangle_{\text{perm}}} \cong 
Y_{*} (Q_{|2L,0\rangle_{\text{perm}}}) \cong
Q_{|L,0\rangle}\oplus Q_{|L-1,0\rangle} \ ,
\end{equation}
where it is understood that $Q_{|L-1,0\rangle}$ is absent when $L=0$.

\subsection{RR-charges}

The interface $_{y}I_{x}$ between the $SU(3)/U(2)$ Kazama-Suzuki model
and the two minimal models can also be used to relate correlators in
these theories. As a simple example we study the RR-charge, which can be
considered as a disc one-point function of the corresponding RR-field. 

\begin{figure}
\centering
\includegraphics[width=\textwidth]{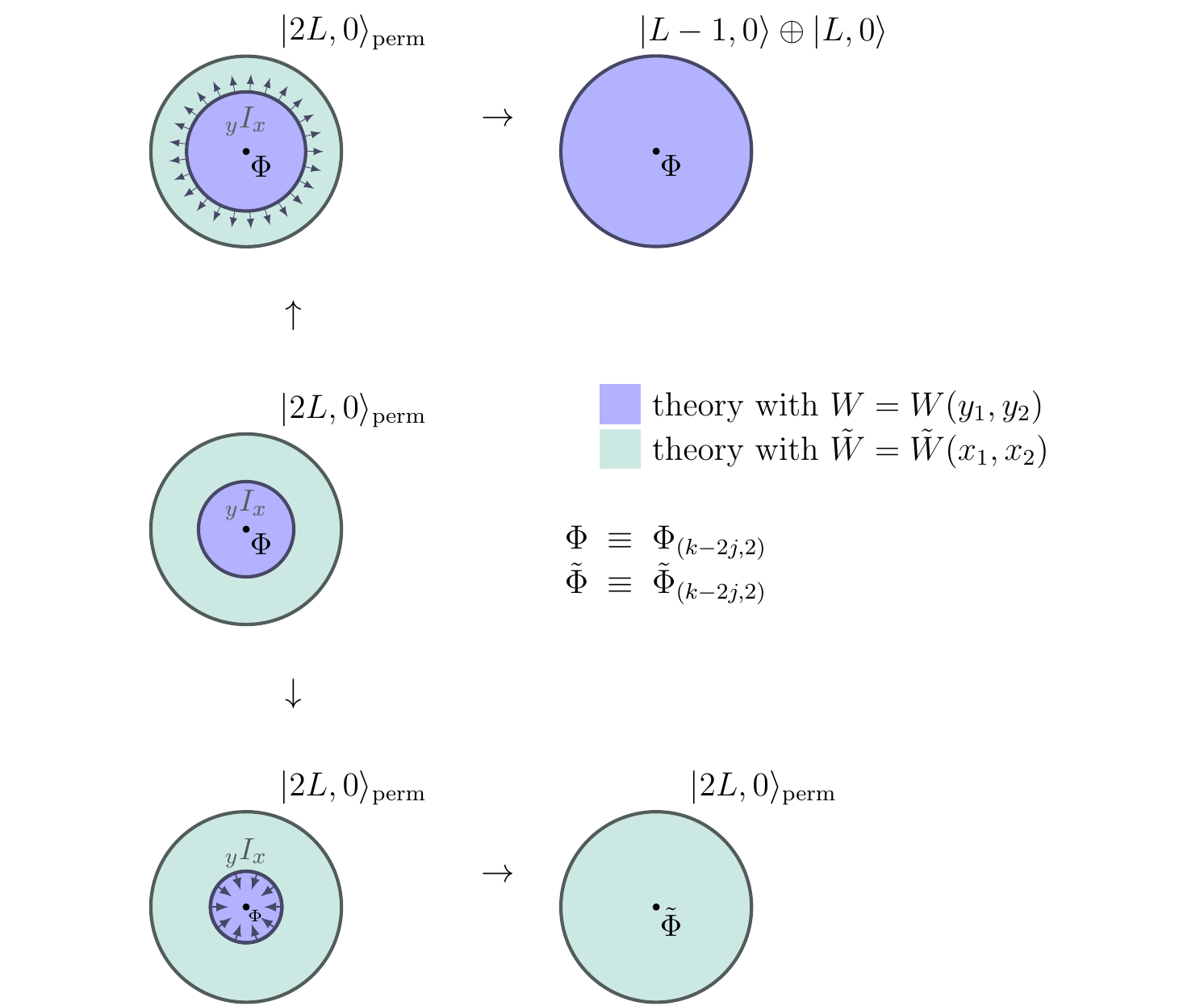}
\caption{\label{fig:interface}Consider a disc correlator with a bulk
field $\Phi$ inserted at the centre, and the interface $_{y}I_{x}$
inserted around it (see the central illustration to the left). Then we
can either shrink the interface around the insertion to produce a
field insertion by a field $\tilde{\Phi}$, or we let the interface
cycle grow until it hits the boundary to produce a new boundary
condition. In this way we can relate two bulk one-point functions on
the disc.}
\end{figure}

The chiral primaries in the $SU(3)/U(2)$ Kazama-Suzuki model are
labelled by an $SU(3)$ representations with Dynkin labels
$(\Lambda_{1},\Lambda_{2})$ (see~\eqref{chiralprimary}) and can be
expressed as polynomials in the variables $y_{1},y_{2}$ (see e.g.\ \cite{Behr:2010ug}),
\begin{equation}
\Phi_{(\Lambda_{1},\Lambda_{2})} (y_{1},y_{2}) =
\sum_{r=0}^{\lfloor\Lambda_{1}/2\rfloor} (-1)^{r}
\binom{\Lambda_{1}-r}{r} y_{1}^{\Lambda_{1}-2r}\,y_{2}^{\Lambda_{2}+r}
\ .
\end{equation}
The chiral primary fields are related to the Ramond ground states by
spectral flow. Only the Ramond ground states with zero
$U(1)_{R}$-charge have non-trivial one-point functions in the
presence of a B-type boundary conditions, the corresponding chiral
primary fields are given by $\Phi_{(k-2j,j)}$ with 
$j=0,\dotsc ,\lfloor\frac{k}{2}\rfloor$. The one-point function in the
presence of the factorisation $Q_{|L,0\rangle}$ is given by
(see~\cite{Behr:2010ug})\footnote{Notice that the expression here
differs from the one in~\cite{Behr:2010ug} by a sign, which is only a
matter of convention regarding the definition of the one-point function.}
\begin{equation}
\langle \Phi_{(k-2j,j)}\rangle_{|L,0\rangle} = -\sum_{i=0}^{L}
\big(\eta^{(2i+1)(j+1)}+\eta^{-(2i+1)(j+1)} \big) \ .
\end{equation}
On the other hand in the minimal models, the chiral primary fields
corresponding to chargeless Ramond ground states are labelled by
\begin{equation}
\Psi_{j} (x_{1},x_{2}) = x_{1}^{j}x_{2}^{k+1-j} \ .
\end{equation}
In the presence of a boundary given by the factorisation
$Q_{|L,M\rangle_{\text{perm}}}$, one can straightforwardly compute the RR one-point
function using the Kapustin-Li formula~\cite{Kapustin:2003ga,Herbst2005}, and one finds
\begin{equation}
\langle \Psi_{j}\rangle_{|L,M\rangle_{\text{perm}}} =
\sum_{i=-\frac{M+L}{2}-1}^{-\frac{M-L}{2}-1} \eta^{(2i+1)(j+1)} \ .
\end{equation}
What is the relation between the RR-charges in the two theories? We
observed before (see~\eqref{KS-bc-from-minmod}) that the interface $_{y}I_{x}$ maps
$|2L,0\rangle_{\text{perm}}$ to $|L,0\rangle \oplus |L-1,0\rangle$. We
therefore have the expectation that\footnote{Similar computations have appeared for (generalised) orbifolds of Landau-Ginzburg models in~\cite{Brunner2014,Brunner2014a}.}
\begin{equation}\label{RRexpectation}
\langle \Phi_{(k-2j,j)}\rangle_{|L,0\rangle \oplus |L-1,0\rangle} = 
\langle \tilde{\Phi}_{(k-2j,j)}\rangle_{|2L,0\rangle_{\text{perm}}} \ ,
\end{equation}
where $\tilde{\Phi}_{(k-2j,j)}$ is the minimal model field that is
obtained when the interface acts on $\Phi_{(k-2j,j)}$. This is
illustrated in figure~\ref{fig:interface}.
In~\cite{Carqueville:2012st,Carqueville2012} it has been worked out how an interface
acts on a bulk field. Applying these methods one can see that the
action of a variable transformation interface $_{x}I_{y}$ on a field
$\Phi(y_{j})$ is in general given by
\begin{equation}
\tilde{\Phi} (x_{i}) = \det \left(\frac{\partial Y_{r}}{\partial
x_{s}} \right) \Phi \big(Y_{j} (x_{i})\big) \ .
\end{equation}
In our case we obtain
\begin{align}
\tilde{\Phi}_{(k-2j,j)} (x_{1},x_{2}) &= (x_{1}-x_{2}) \cdot \Phi_{(k-2j,j)} (x_{1}+x_{2},x_{1}x_{2})\nonumber\\
&= (x_{1}-x_{2}) \sum_{i=0}^{k-2j} x_{1}^{i+j}x_{2}^{k-i-j}\nonumber\\
&= x_{1}^{k-j+1}x_{2}^{j} - x_{1}^{j}x_{2}^{k-j+1} \nonumber\\
&= \Psi_{k-j+1} (x_{1},x_{2}) - \Psi_{j} (x_{1},x_{2})\ .
\end{align}
Therefore the right hand side of~\eqref{RRexpectation} evaluates to
\begin{align}
\langle \tilde{\Phi}_{(k-2j,j)}\rangle_{|2L,0\rangle_{\text{perm}}} &=
\sum_{i=-L-1}^{L-1}\big(\eta^{(2i+1)(k+2-j)}-\eta^{(2i+1)(j+1)}
\big)\nonumber\\
&=- \sum_{i=-L-1}^{L-1}\big(\eta^{-(2i+1)(j+1)}+\eta^{(2i+1)(j+1)}
\big)\nonumber\\
&=- \sum_{i=0}^{L-1} \big(\eta^{(2i+1)(j+1)} + \eta^{-(2i+1)(j+1)}
\big) \nonumber\\
&\qquad - \sum_{i=0}^{L} \big(\eta^{(2i+1)(j+1)} + \eta^{-(2i+1)(j+1)} \big)
\ ,
\end{align}
which precisely equals the left hand side
of~\eqref{RRexpectation}. 

\subsection{Higher factorisations from cones}

To construct matrix factorisations for other rational boundary conditions,
one can make use of the known flows between different boundary
states~\cite{Behr:2010ug}, which we will review now. Evaluating the
flow~\eqref{simplestflow} for $\ell =0$,
we obtain
\begin{equation}\label{flow}
|L,0\rangle + |L,1\rangle \longrightarrow |L-1,0\rangle +
|L,0\rangle + |L+1,0\rangle \ .
\end{equation}
Translated in the matrix factorisation language this means that we
expect that the factorisation corresponding to the right hand side can
be obtained as a cone from the two factorisations that correspond to
the left hand side of the flow~\eqref{flow}. 
In other words,
$Q_{|L,1\rangle}$ can be obtained as a cone from
$Q_{\overline{|L,0\rangle}}$ and the superposition $Q_{|L-1,0\rangle}\oplus
Q_{|L,0\rangle}\oplus Q_{|L+1,0\rangle}$. This in turn can be
rewritten~\cite{Behr:2010ug} as a cone of $Q_{|L,0\rangle}$ and the factorisation
\begin{equation}
\tilde{Q}_{L} = \begin{pmatrix}
0 & \cJ_{L+1}\cJ_{|L-1,0\rangle} \\
\cJ_{L}\cJ_{\overline{|L+1,0\rangle}} & 0
\end{pmatrix} \ .
\end{equation}
Explicitly we find
\begin{equation}
Q_{|L,1\rangle} = C 
\left( 
Q_{|L,0\rangle},\tilde{Q}_{L}, y_{1}
\begin{pmatrix}
0 & \cJ_{|L-1,0\rangle} \\
-\cJ_{\overline{|L+1,0\rangle}} & 0 
\end{pmatrix} 
\right) \ ,
\end{equation}
such that
\begin{subequations}
\begin{align}\label{L1factorisation1}
Q^{(1)}_{|L,1\rangle} &= 
\begin{pmatrix}
\cJ_{L} & 0\\
y_{1} & \cJ_{L+1}
\end{pmatrix}\cJ_{|L-1,0\rangle}\\
Q^{(0)}_{|L,1\rangle} &=
\begin{pmatrix}
\cJ_{L+1} & 0\\
-y_{1} & \cJ_{L}
\end{pmatrix}\cJ_{\overline{|L+1,0\rangle}} \ .
\end{align}
\end{subequations}
In principle one can use the same strategy to obtain factorisations
for $|L,\ell\rangle$ for $\ell\geq 2$ by the cone construction. There are,
however, two obstacles in this approach, one of technical nature, the
other one being a conceptual problem. On the technical side one faces
the problem that the factorisations in question become larger and
larger, and the computations are feasible only by means of a computer
program. In fact, the flow rule~\eqref{flow} leads to a realisation of
$Q_{|L,1\rangle}$ as an $8\times 8$ matrix (that can then be reduced
to the $4\times 4$ matrix that we saw above), and similarly the
general flow rule~\eqref{simplestflow} leads to an ansatz where the
$Q_{|L,2\rangle}$ factorisations are already $32\times 32$ matrices,
and the $Q_{|L,3\rangle}$ are of size $128\times 128$.
Even with the help of rather efficient SINGULAR
codes and considerable amounts of computer processing power, the
authors were not able to push this type of search much beyond the
$Q_{|L,2\rangle}$ type factorisations, with only a few sporadic matches for
$Q_{|L,3\rangle}$, and the codes not being executable due to
memory limitations already for the $Q_{|L,4\rangle}$ type factorisations.

There is also a conceptual problem in this approach. For the
$|L,1\rangle$ boundary states, one can uniquely identify the field
that is responsible for the flow by its $U(1)_{R}$-charge, and
therefore one is led to a unique ansatz for the cone. This is in general
not true for $|L,2\rangle$ and beyond. This problem is also reflected
by the presence of marginal boundary fields for the $|L,2\rangle$
boundary condition (if $L\not= k/2$): it can be smoothly deformed to other
boundary states. Correspondingly, the associated matrix factorisations
can be deformed, and within this continuous family of $|L,2\rangle$-like
factorisations it is hard to identify the one that corresponds
precisely to $|L,2\rangle$.

This is why we look for a different approach to obtain the higher
factorisations, which will be based on special operator-like
defects in the theory as we will discuss in the following.

\subsection{Higher factorisations from defect fusion}\label{sec:HF1}

Besides the cone construction, which we employed in the last
subsection, we can also use fusion of defects or interfaces to
generate new factorisations. We have seen in
section~\ref{sec:polynomialMF} that we can generate the $|L,0\rangle$
factorisations from permutation factorisations in minimal models by
fusing the variable transformation interface $_{y}I_{x}$, namely
\begin{equation}
_{y}I_{x}\otimes Q_{|2L,0\rangle_{\text{perm}}} \cong  Y_{*} (Q_{|2L,0\rangle_{\text{perm}}}) \cong Q_{|L-1,0\rangle}
\oplus Q_{|L,0\rangle} \ .
\end{equation} 
What happens if we tensor $_{y}I_{x}$ to other permutation
factorisations? Let us look at the factorisations corresponding to
the permutation boundary states
$|2L+1,1\rangle_{\text{perm}}$. From~\eqref{permutationMF} we see that
the upper right entry is 
\begin{equation}
Q^{(1)}_{|2L+1,1\rangle_{\text{perm}}} = (x_{1}-\eta^{-2L-3}x_{2}) (x_{1}-\eta^{-2L-1}x_{2})
\prod_{j=0}^{L-1}\cJ_{j} (y_{1},y_{2})\Big|_{\ytox}\ .
\end{equation}
Fusing the variable transformation interface to this factorisation,
i.e.\ applying the functor $Y_{*}$, we obtain
\begin{align}
&Y_{*}\big( Q^{(1)}_{|2L+1,1\rangle_{\text{perm}}} \big) 
= \prod_{j=0}^{L-1} \cJ_{j} (y_{1},y_{2})\times \nonumber\\
&\times \begin{pmatrix}
\frac{1+\eta^{-4L-4}}{2}y_{1}^{2} - (1-\eta^{-2L-1}) (1-\eta^{-2L-3})
& \frac{1-\eta^{-2L-4}}{2}y_{1} (y_{1}^{2}-4y_{2})  \\
\frac{1-\eta^{-2L-4}}{2}y_{1} & \frac{1+\eta^{-4L-4}}{2}y_{1}^{2} - (1-\eta^{-2L-1}) (1-\eta^{-2L-3})
\end{pmatrix}\nonumber\\
& \to \begin{pmatrix}
\cJ_{L} & 0\\
y_{1} & \cJ_{L+1}
\end{pmatrix}\cJ_{|L-1,0\rangle}\ ,
\end{align}
where we performed a similarity transformation in the last step. This
is precisely $Q^{(1)}_{|L,1\rangle}$ (see~\eqref{L1factorisation1}),
so that we find
\begin{equation}
Y_{*}\big(Q_{|2L+1,1\rangle_{\text{perm}}} \big) \cong
Q_{|L,1\rangle} \ .
\end{equation}
We found again that a rational boundary condition is mapped to a
rational one by the variable transformation interface.

There is, however, much more that we can conclude from this
finding. In fact we expect from the rational conformal field theory description that
there is a defect $D_{[(0,0),0;1,3]}$ that maps $|L,0\rangle$ to
$|L,1\rangle$. A natural ansatz would be to look for a rational defect $\tilde{D}$
in the minimal model theory, and then fusing it from the left with
$_{y}I_{x}$ and from the right with $_{x}I_{y}$ to obtain a defect in
the Kazama-Suzuki model,
\begin{equation}
D_{[(0,0),0;1,3]} = {}_{y}I_{x}\otimes \tilde{D}\otimes {}_{x}I_{y} \ .
\end{equation}
We know that under fusion with $D_{[(0,0),0;1,3]}$ the factorisation
$Q_{|L,0\rangle}$ should be mapped to $Q_{|L,1\rangle}$. Fusing the
variable transformation interface onto $Q_{|L,0\rangle}$ leads to
the factorisation $Q_{|2L+1,-1\rangle_{\text{perm}}}$. On the other hand we just derived
that $Q_{|2L+1,1\rangle_{\text{perm}}}$ is mapped to $Q_{|L,1\rangle}$
when we fuse $_{y}I_{x}$. Therefore we demand that the defect
$\tilde{D}$ maps $|2L+1,-1\rangle_{\text{perm}}$ to $|2L+1,1\rangle_{\text{perm}}$.
In fact there is a symmetry defect, $Q_{\{1\}}\otimes
Q_{\{\eta^{2}\}}$ that acts as the identity defect in the first
minimal model factor, and as the symmetry defect realising the
automorphism $\sigma_{\eta^{2}}:x_{2}\to\eta^2 x_{2}$ in the second
minimal model. We therefore conjecture that
\begin{equation}
\tilde{D} = Q_{\{1\}} \otimes Q_{\{\eta^{2}\}} \ .
\end{equation}
This is again a simple example of a variable transformation interface,
whose fusion is described by the functor $\sigma_{\eta^{2}}^{*}$ that
acts trivially on the variable $x_1$ and replaces the variable $x_{2}$ by $\eta^{2}x_{2}$. The fusion of
the defect $D_{[(0,0),0;1,3]}$ can then be described by the functor
\begin{equation}\label{defofD}
D_{(1)} = Y_{*} \circ \sigma_{\eta^{2}}^{*} \circ Y^{*} \ .
\end{equation}
We have thus identified a candidate for a defect in the Landau-Ginzburg
theory whose action on the boundary conditions $Q_{|L,0\rangle}$
coincides precisely with what we expect from the fusion of the defect
$D_{[(0,0),0;1,3]}$ on the boundary condition $|L,0\rangle$. This is
of course not a proof that we identified the defect correctly in the
Landau-Ginzburg model, and we briefly want to discuss two obvious ways
how one could try to modify the proposal. Firstly we might modify the proposal by
choosing instead of $Q_{\{1 \}}\otimes Q_{\eta^{2}}$ the symmetry defect 
\begin{equation}\label{}
Q_{\{\eta^{2m}\}}\otimes Q_{\{\eta^{2(m+1)}\}} \ ,
\end{equation}
which would lead to the same action on boundary conditions
$Q_{|L,0\rangle}$. To decide which choice is the correct one, we have
to act with the defect on other defects. From the conformal field theory we
expect the fusion
\begin{equation}
D_{[(0,0),0;1,3]} * D_{[(0,0),0;1,3]} = D_{[(0,0),0;2,6]} \oplus
D_{[(0,0),0;0,6]} \ .
\end{equation}
The second defect is a symmetry defect that corresponds to the phase
shifts
\begin{equation}
y_{1}\mapsto \eta^{2}y_{1}\,,\; y_{2}\mapsto \eta^{4}y_{2}\ ,
\end{equation}
which means that we know its identification on the Landau-Ginzburg
side. By looking at the above fusion of the defect $D_{[(0,0),0;1,3]}$
with itself in the Landau-Ginzburg theory (which we will present in~\cite{NBSFtwo2014}), we can therefore confirm that we made the correct choice.

The second obvious question one should investigate is whether there
are any smooth deformations of this defect, so that there would be a
whole family of defects with similar properties. As one can show
from a computation of the conformal field theory spectrum, we do not expect any fermionic morphisms of
the corresponding matrix factorisations, and 
therefore no deformations. This provides
further evidence that we have identified the defect correctly.

Having identified $D_{[(0,0),0;1,3]}$ in the Landau-Ginzburg model,
one can then use it to construct higher factorisations, which we will
do in the following subsection.  

\subsection{Matrix factorisations for all rational boundary conditions}

With the help of the defect $D_{[(0,0),0;1,3]}$ we can in principle
determine all matrix factorisations corresponding to rational boundary
conditions. In fact, we know from the conformal field theory that
(see~\eqref{examplefusion})
\begin{equation}
D_{[(0,0),0;1,3]}  |L,\ell \rangle  = |L,\ell -1\rangle 
+ |L,\ell+1\rangle \ , 
\end{equation}
where it is understood that the first boundary condition on the right
is not present for $\ell=0$. For the factorisations this means that
\begin{equation}
D_{(1)} \big(Q_{|L,\ell\rangle} \big) \cong Q_{|L,\ell-1\rangle}
\oplus Q_{|L,\ell+1\rangle} \ .
\end{equation}
Starting from $Q_{|L,0\rangle}$ one can generate all
$Q_{|L,\ell\rangle}$ by successively applying $D_{(1)}$. The
technical challenge that remains is to decompose the fusion result
into the direct sum of two factorisations.

\subsubsection{A closed formula for rational matrix factorisations
$Q_{|0,\ell\rangle}$}\label{sec:zeroell}

We now want to investigate this problem for the factorisations of type $Q_{|0,\ell\rangle}$. They are
generated from the factorisation $Q_{|0,0\rangle}$ which is a $2\times
2$ matrix whose upper right block $Q_{|0,0\rangle}^{(1)}$ is the polynomial $\cJ_{0}$
(see~\eqref{eq:QLzeroMFs}). Applying $D_{(1)}$ once we obtain a matrix
factorisation for $|0,1\rangle$ whose upper right block (after a
similarity transformation) is given by (see~\eqref{L1factorisation1})
\begin{equation}
D_{(1)} (\cJ_{0}) \cong  Q_{|0,1\rangle}^{(1)} = \begin{pmatrix}
\cJ_{0} & 0\\
y_{1} & \cJ_{1}
\end{pmatrix} \ .
\end{equation}
We see the polynomial factors $\cJ_{n}$ appearing on the diagonal. In
the full matrix factorisation $Q_{|0,1\rangle}$ they appear as part of
the matrix factorisation blocks 
\begin{equation}
Q_{n} = \begin{pmatrix}
0 & \cJ_{n}\\
\bar{\cJ}_{n} & 0
\end{pmatrix}
\end{equation}
with $\bar{\cJ}_{n}=W^{y}_{2;k}/\cJ_{n}$.

When we want to apply $D_{(1)}$ once more, we first have to understand its
action on these blocks $Q_{n}$. We will need later a result not only
for $Q_{0}$ and $Q_{1}$, but for a general factorisation $Q_{n}$. Introducing the notation
\begin{align}\label{defplusminus}
\plus_{p}&=\frac{1}{2} \big(1+\eta^{p}\big) & \minus_{p}&=\frac{1}{2}\big (1-\eta^{p}\big) \ ,
\end{align}
the factor $\cJ_{n}$ (see~\eqref{cJs}) can be expressed as
\begin{equation}\label{eq:JpolyDef}
\cJ_{n} = y_{1}^{2}\minus_{2n+1}\minus_{-2n-1} +
\lambda_{1}^{2}\plus_{2n+1}\plus_{-2n-1} \ ,
\end{equation}
where 
\begin{equation}
\lambda_{1}^{2}:= y_{1}^{2}-4y_{2} = (x_{1}-x_{2})^{2}\Big|_{\xtoy} \ .
\end{equation}
Applying $D_{(1)}$ (given in~\eqref{defofD}) to $Q_{n}$ we find for
the upper right block $Q_{n}^{(1)}=\cJ_{n}$
\begin{align}
D_{(1)} \big(\cJ_{n} \big) &= Y_{*} \Big(  (x_{1}+\eta^{2}x_{2})^{2}\minus_{2n+1}\minus_{-2n-1} +
(x_{1}-\eta^{2}x_{2})^{2 }\plus_{2n+1}\plus_{-2n-1} \Big)\\
&= \begin{pmatrix}
y_1^2\minus_{2n+3}\minus_{-2n+1}+\lambda_1^2\plus_{2n+3}\plus_{-2n+1} & 2y_1\lambda_1^2 \minus_{2}\plus_{2}\\
2y_1\minus_{2}\plus_{2}
&y_1^2\minus_{2n+3}\minus_{-2n+1}+\lambda_1^2\plus_{2n+3}\plus_{-2n+1}
\end{pmatrix} \\ 
&= (\cU^{(0)}_n)^{-1}\cdot \cJ_{n(1)}\cdot \cU^{(1)}_n 
\label{eq:D1onJn}
\end{align}
with
\begin{equation}\label{cJn1}
\cJ_{n(1)}=\begin{pmatrix}
\cJ_{n-1} & 0\\
y_1 & \cJ_{n+1}
\end{pmatrix}\ .
\end{equation}
In the last step we performed a similarity transformation to define a
convenient form $Q_{n(1)}$ for the factorisation $D_{(1)}(Q_{n})$,
\begin{equation}\label{Qn1}
Q_{n(1)} = \cU_{n} \cdot \big(D_{(1)} (Q_{n}) \big)\cdot
\big(\cU_n\big)^{-1}\  ,    
\end{equation}
where the transformation $\cU_{n}:={}^{1}\cU_{n}$is defined by
\begin{align}
{}^{r}\cU_n&:=
\cU_{row\times}\left(r+1;\frac{1}{\minus_{4}}\right)\cdot
\cU_{row\times}\left(r;\frac{\plus_{2n-1}}{\plus_{2n+3}}\right)\cdot
\cU_{col\times}\left(r+1;\minus_{4}\frac{\plus_{-2n-3}}{\plus_{-2n+1}}\right)\cdot\nonumber\\
&\qquad \cdot
\cU_{col}\left(r,r+1;-\frac{y_1 \minus_{2n+3}}{\plus_{2n+3}}\right)\cdot
\cU_{row}\left(r+1,r;\frac{y_1 \minus_{2n-1}}{\plus_{2n-1}}\right)\ .
\label{cUr}
\end{align}
Here, $\cU_{row\times}(r;\alpha)$ ($\cU_{col\times}(r;\alpha)$) has the
effect of multiplying row $r$ (column $r$) of the upper right block
$Q^{(1)}$ of a matrix factorisation with the constant
$\alpha$. $\cU_{col}(r,s;\alpha)$ ($\cU_{row} (r,s;\alpha)$) has the
effect on the block $Q^{(1)}$ of adding row $r$ (column $r$) multiplied by
$\alpha$ to row $s$ (column $s$). The precise conventions and explicit formulae for the similarity transformations
are summarised in Appendix~\ref{sec:Utrafos}.

Let us now apply $D_{(1)}$ on $Q_{n(1)}$. The upper right block
$Q_{n(1)}^{(1)}=\cJ_{n(1)}$ is given in~\eqref{cJn1}, and it has the
factors $\cJ_{n-1}$ and $\cJ_{n+1}$ on the diagonal, which will be
mapped to $D_{(1)}(\cJ_{n\pm 1})$. We then directly apply the
similarity transformations to bring those to the form $\cJ_{n\pm 1(1)}$,
\begin{align}
D_{(1)} (\cJ_{n(1)}) & =\begin{pmatrix}
D_{(1)} (\cJ_{n-1}) & 0\\
D_{(1)} (y_{1}) & D_{(1)} (\cJ_{n+1})
\end{pmatrix}\\
&= \begin{pmatrix}
\cU_{n-1}^{(0)} & 0\\
0 & \cU_{n+1}^{(0)}
\end{pmatrix}^{\!\!\!-1}\cdot
\begin{pmatrix}
\cJ_{{n-1}(1)} & 0\\
\widetilde{D_{(1)}(y_1)}_n & \cJ_{{n+1}(1)}
\end{pmatrix}\cdot \begin{pmatrix}
\cU_{n-1}^{(1)} & 0\\
0 & \cU_{n+1}^{(1)}
\end{pmatrix}
\end{align}
with
\begin{align}
\widetilde{D_{(1)}(y_1)}_n&= \cU_{n+1}^{(0)} \cdot
D_{(1)}(y_{1}) \cdot (\cU_{n-1}^{(1)})^{-1}\\
&=\begin{pmatrix}
\frac{y_1\plus_{2n+3}}{\plus_{2n+5}} & \frac{\cJ_n \minus_{2}\minus_{4}}{\plus_{2n+5}\plus_{-2n+3}}\\
\frac{1}{2\plus_{2}} & \frac{y_1\plus_{-2n+1}}{\plus_{-2n+3}}
\end{pmatrix}\ .
\end{align}
The effect of the similarity transformation is summarised in the
transformation
\begin{equation}
\cU_{n(1)}^{a}:={}^3{}\cU_{n+1}\cdot {}^1{}\cU_{n-1} \ ,
\end{equation}
where the left superscript $j$ on ${}^{j}\cU_{m}$ denotes the row and
column where the corresponding $2\times 2$-block $\cU_{m}$ starts (in accordance
with the definition in~\eqref{cUr}).

We can perform further similarity transformations to bring
$D_{(1)}(Q_{n(1)})$ into a convenient form: 
\begin{equation}\label{firstsplitting}
\begin{split}
D_{(1)}\big( \cJ_{n(1)}\big)
&= \big(\cU_{n(1)}^{b(0)} \cdot \cU_{n(1)}^{a(0)} \big)^{-1}\cdot \begin{pmatrix}
\cJ_{n-2} & 0 & 0 & 0\\
y_1 & \cJ_{n} & 0 & 0\\
0 & -\frac{\cJ_n}{2\plus_{2}\chi_{(n)}} & \cJ_n & 0\\
\frac{1}{2\plus_{2}} & 0 & y_1 & \cJ_{n+2}
\end{pmatrix}\cdot \cU_{n(1)}^{b(1)}\cdot\cU_{n(1)}^{a(1)}\\
&= \big(\cU_{n(1)}^{c(0)}\cdot \cU_{n(1)}^{b(0)}\cdot \cU_{n(1)}^{a(0)}\big)^{-1}\cdot\begin{pmatrix}
\cJ_{n-2} & 0 & 0 & 0\\
y_1 & 0 & \cJ_n & 0\\
0 & \cJ_n & 0 & 0\\
\chi_{(n)} & 0 & y_1 & \cJ_{n+2}
\end{pmatrix}\cdot\cU_{n(1)}^{c(1)}\cdot\cU_{n(1)}^{b(1)}\cdot\cU_{n(1)}^{a(1)}\ .
\end{split}
\end{equation}
Here, the transformation $\cU_{n(1)}^{b}$ is a simple row and column operation that deletes the entries $\propto y_1$,
\begin{equation}\label{defcUb}
\cU_{n(1)}^b={}^2{}\cU_{n(1)}\,,\quad {}^r{}\cU_{n(1)}:=
\cU_{col}\left(r+1,r;-\frac{\plus_{-2n+1}}{\plus_{-2n+3}}\right)\cdot\cU_{row}\left(r,r+1;-\frac{\plus_{2n+3}}{\plus_{2n+5}}\right)\ ,
\end{equation}
while the transformation $\cU_{n(1)}^c$ is defined as 
\begin{equation}\label{defcUc}
\begin{split}
\cU_{n(1)}^c&={}_{\tilde{c}}^2{}\cU_{n(1)}\cdot{}_c^2{}\cU_{n(1)}\,\\
{}_c^r{}\cU_{n(1)}&:=\cU_{col}\left(r,r+1;2\plus_{2}\chi_{(n)}\right)\cdot\cU_{row}\left(r+1,r;2\plus_{2} \chi_{(n)}\right)\\
{}_{\tilde{c}}^r{}\cU_{n(1)}&:=
\cU_{row\times}\left(r-1;\frac{1}{2\plus_{2}\chi_{(n)}}\right)\cdot
\cU_{col\times}\left(r-1;2\plus_{2}\chi_{(n)}\right)\\
&\qquad \cdot \cU_{row\times}\left(r;\frac{1}{2\plus_{2}\chi_{(n)}}\right)\cdot 
\cU_{col\times}\left(r;-2\plus_{2}\chi_{(n)}\right) \ .
\end{split}
\end{equation}
For convenience we introduced the quantities
\begin{equation}\label{defchi}
\chi_{(p)}:= \frac{\plus_{-2p+3}\, \plus_{2p+5}}{4\plus_{2}^{2}\,\plus_{-2p+1}\, \plus_{2p+3}}\ .
\end{equation}
Looking at~\eqref{firstsplitting} we see that the matrix
factorisations can be split into the factorisation $Q_{n}$ and a new
factorisation $Q_{n(2)}$ whose upper right block 
$Q_{n(2)}^{(1)}=\cJ_{n(2)}$ is
\begin{equation}
\cJ_{n(2)} = \begin{pmatrix}
\cJ_{n-2} & 0 & 0 \\
y_{1} & \cJ_{n} & 0 \\
\chi_{(n)} & y_{1} & \cJ_{n+2}
\end{pmatrix} \ .
\end{equation}
In particular, we can identify the factorisation for the boundary state
$|0,2\rangle$ as $Q_{|0,2\rangle}^{(1)}=\cJ_{0(2)}$. 

One can now go on and apply $D_{(1)}$ again. We will show in appendix~\ref{app:B} that in this way one generates a family of factorisations $Q_{n(m)}$ with the property
\begin{equation}
D_{(1)} (Q_{n(m)}) \cong Q_{n(m-1)} \oplus Q_{n(m+1)} \ .
\end{equation}
The upper right block $\cJ_{n(m)}\equiv Q^{(1)}_{n(m)}$ of $Q_{n(m)}$ is given by
\begin{equation}
\begin{split}
\cJ_{n(m)}&=\begin{pmatrix}
\cJ_{n-m} & 0 & \cdots &&&&&&\\
y_1 & \cJ_{n-m+2} & 0 & \cdots &&&&&&\\
\chi_{(n-m+2)} & y_1 & \cJ_{n-m+4} & 0 & \cdots &&&&&\\
0 & \chi_{(n-m+4)} & y_1 & \cJ_{n-m+6} & 0 & \cdots &&&&\\
\vdots & \ddots & \ddots & \ddots & \ddots & \ddots & &&&\\
\\
&&&&&& 0 & \chi_{(n+m-2)} & y_1 & \cJ_{n+m}
\end{pmatrix}\ .
\end{split}\label{eq:rMF}
\end{equation}
This formula applies for odd level $k$ for all $m\leq k+2$, whereas
for even level $k$ it applies for $m+|n|\leq k/2$. 

In particular one therefore has found matrix factorisations
$Q_{|0,\ell\rangle}=Q_{0(\ell)}$ for the rational boundary states
$|0,\ell\rangle$. For odd $k$ this covers all boundary states of this
type, whereas for even $k$ we have the restriction $\ell \leq k/2$.
Note however that (see~\eqref{bsidentification})
\begin{equation}
|L,\ell \rangle  = \overline{|L,k+1-\ell \rangle} \ ,
\end{equation}
therefore boundary states with $\ell\geq k/2+1$ can be related to boundary
states with smaller label. The operation of taking the anti-boundary state corresponds in the
matrix factorisation to an exchange of the blocks $Q^{(0)}$ and
$Q^{(1)}$. Therefore we have found factorisations for all boundary
states of the form $|0,\ell\rangle$.
\smallskip

As presented in appendix~\ref{sec:closed}, it is possible to
find a very compact alternative closed expression for the form of both
the $\cJ_{n(m)}\equiv Q^{(1)}_{n(m)}$ block as well as of the $\cE_{n(m)}\equiv Q^{(0)}_{n(m)}$ block of the
matrix factorisations $Q_{n(m)}$. Referring to the appendix for the computational details, we would just like to mention here that the derivation is based on two major steps. In the first step, the structure of the $\cE_{n(m)}$ blocks is inductively derived from the explicit formula~\eqref{eq:rMF} for $\cJ_{n(m)}$ via the basic equation
\begin{equation}
Q_{n(m)}^2=W\cdot \mathbb{1}\quad \Leftrightarrow\quad
\cE_{n(m)}=W\cdot\left(\cJ_{n(m)}\right)^{-1}\ .
\end{equation}
The second step consists in applying a series of row and column
operations on the $\cJ_{n(m)}$ block in order to ``clear out'' all
rows and columns that intersect at a constant entry. According
to~\eqref{eq:rMF}, this leaves a $2\times2$ non-trivial block
$\widehat{\cJ}_{n(m)}$ in direct sum with $m-1$ trivial matrix factorisation blocks $\cJ_{triv}$. Upon closer inspection, the aforementioned similarity transformations induce operations on the $\cE_{n(m)}$ block that leave the $2\times2$ subblock $\widehat{\cE}_{n(m)}$ formed from the overlap of the last two lines and the first two columns of $\cE_{n(m)}$ invariant. But since $\widehat{\cE}_{n(m)}$ is thus just a subblock of $\cE_{n(m)}$, in contrast to $\widehat{J}_{n(m)}$ we already know an explicit formula for $\widehat{\cE}_{n(m)}$, and thus in turn also for $\widehat{\cJ}_{n(m)}$:
\begin{equation}\label{twotworealisation}
\boxed{
\begin{aligned}
\widehat{\cE}_{n(m)}&=\begin{pmatrix}
\Psi_{n-1(m-1)} & \Psi_{n(m-2)}\\
\Psi_{n(m)} & \Psi_{n+1(m-1)}
\end{pmatrix}\ ,\\
\widehat{\cJ}_{n(m)}&=W\widehat{\cE}_{n(m)}^{-1}=
\frac{1}{W}
\frac{\prod_{j=0}^{m}\cJ_{n-m+2j}}{\prod_{j=1}^{m-1}\chi_{(n-m+2j)}}
\begin{pmatrix}
\Psi_{n+1(m-1)} & -\Psi_{n(m-2)}\\
-\Psi_{n(m)} & \Psi_{n-1(m-1)}
\end{pmatrix}\ .
\end{aligned}}
\end{equation}
The explicit formula for the entries $\Psi_{n(m)}$ is given
in~\eqref{app:solforPsi} in appendix~\ref{sec:closed}.

\subsubsection{A closed formula for all rational matrix factorisations}

To obtain expressions for all rational matrix factorisations, we start from the factorisations
$Q_{|L,0\rangle}$ and apply $D_{(1)}$ successively to generate
factorisations for the boundary states $|L,\ell\rangle$,
\begin{equation}
D_{(1)}\big(Q_{|L,\ell\rangle} \big) \cong Q_{|L,\ell-1\rangle} \oplus
Q_{|L,\ell+1\rangle} \ .
\end{equation}
The biggest computational problem is then the decomposition into the
elementary factorisations on the right hand side. This was already
tedious for $L=0$ where we started from a degree~2 polynomial
$\cJ_{0}$, so a priori, it appears hopeless to find a
closed formula for the factorisations $Q_{|L,\ell\rangle}$ with $\ell>1$, where
the starting polynomial 
\begin{equation}
\cJ_{|L,0\rangle} = \prod_{i=0}^L \mathcal{J}_i
\end{equation}
is of degree $2(L+1)$. We may however rewrite the higher polynomial
factorisations as cones of the elementary polynomial factorisations
(see e.g.\ \cite{Behr:2010ug}), such that (we will again only write
the upper right block of the matrix factorisations)
\begin{equation}\label{eq:JLzeroAlt}
\mathcal{J}_{|L,0\rangle}=\prod_{i=0}^L \mathcal{J}_i \cong \begin{pmatrix}
\mathcal{J}_0 & 0 & \cdots &&&&\\
1 & \mathcal{J}_1 & 0 & \cdots &&&\\
0 & 1 & \mathcal{J}_2 & 0 & \cdots & &\\
\vdots & \ddots & \ddots & \ddots & &&\\
&&&0& 1& \mathcal{J}_{L-1} & 0\\
&&&& 0& 1 & \mathcal{J}_L
\end{pmatrix}\ .
\end{equation}
Each of the diagonal entries of the cone is simply a polynomial
factor $\cJ_{n}$ of degree $2$. When we now apply $D_{(1)}$
successively, we can in principle use our results of the previous
subsection to obtain factorisations with blocks $\cJ_{n(m)}$ on the
diagonal. 

The difficulty in this approach is that the similarity transformations
that are used to arrive at the blocks $\cJ_{n(m)}$ will also affect
the morphisms. When we apply $D_{(1)}$ in the first step its action on
the morphisms $1$ is trivial,
\begin{equation}
D_{(1)} (1) = \mathbb{1} \ ,
\end{equation}
but the similarity transformations will produce non-trivial entries. As an example
consider the matrix
\begin{equation}
\cJ_{p,q} = \begin{pmatrix}
\cJ_{p} & 0\\
1 & \cJ_{q}
\end{pmatrix} \ .
\end{equation}
When we apply $D_{(1)}$ on it and transform the diagonal blocks
$D_{(1)}\big(\cJ_{n}\big)$ into the form $\cJ_{n(1)}$
(see~\eqref{cJn1}) via the similarity transformations $\cU_{n}$
(see~\eqref{Qn1}), we obtain
\begin{equation}
D_{(1)}\big( \mathcal{J}_{p,q}\big) \cong \begin{pmatrix}
\mathcal{J}_{p-1} & 0 & 0 & 0 \\
y_1 & \mathcal{J}_{p+1} & 0 & 0\\
\frac{\plus_{2q-1}}{\plus_{2q+3}} & \frac{y_1\minus_{4}\minus_{2q-2p-4}}{\plus_{2q+3}\plus_{-2p+1}} & \mathcal{J}_{q-1} & 0\\
0 & \frac{\plus_{-2p-3}}{\plus_{-2p+1}} & y_1 & \mathcal{J}_{q+1}\\
\end{pmatrix}\ .
\end{equation}
While now the diagonal blocks are in the right form to apply our
inductive mechanism for finding the result of applying $D_{(1)}$ to
them, we observe that since now the morphisms between the
$Q_{n(1)}$-type blocks have an entry of polynomial degree $>0$
($\propto y_1$), each time we apply $D_{(1)}$ we will generate
consecutively higher degree polynomial morphism entries, thus leading
to an extremely complex morphism structure.

We have instead to look for an alternative standard form for the
$\cJ_{n(m)}$ that is obtained by using similarity transformations
that leave the morphisms (the identity matrices) unchanged.
A prototype of such a transformation is one that 
\begin{itemize}
\item does not depend on $n$, and
\item has identical diagonal blocks, $\cU^{(0)}=\cU^{(1)}$.
\end{itemize}
Then the morphism entries are unaffected,
\begin{equation}
\cU^{(0)}\cdot \mathbb{1}\cdot \big(\cU^{(1)}\big)^{-1} = \mathbb{1} \ .
\end{equation}
Our strategy, however, was to allow for all similarity transformations
a priori, and then make sure at the end that all morphisms are again
identity matrices. We conjecture that it is enough to use
transformations with the two properties described above, but it is not
guaranteed from our analysis.

To describe the alternative standard form we found in this way, we
have to introduce some notation. First let us define a 
generalisation of the functor $D_{(1)}$,
\begin{equation}
\tilde{D}_{0,m} := Y_{*} \circ \sigma^{*}_{\eta^{2m}}\circ Y^{*} \ ,
\end{equation}
i.e.\ we first express the variables $y_{i}$ through the $x_{j}$, then
map $x_{2}\mapsto \eta^{2m}x_{2}$, and then apply the functor $Y_{*}$
to again obtain a matrix in the variables $y_{i}$. For $m=1$ we have
$D_{(1)}=\tilde{D}_{0,1}$ (see~\eqref{defofD}). The action of
$\tilde{D}_{0,m}$ on an elementary polynomial factor $\cJ_{n}$ is
given by\footnote{Note that the entries of $\Lambda$ as well as
$\lambda_{1}$ are not elements of the polynomial ring $\bC[y_1,y_2]$,
but that the combination $\lambda_{1}\Lambda$ that appears in the
formulae has entries that can be written as polynomials in $y_{1},y_{2}$.}
\begin{equation}
\begin{split}\label{eq:D0monJn}
\tilde{D}_{0,m}\big(  \mathcal{J}_n\big) 
&=\left(y_1^2 \minus_{2m+2n+1}\minus_{2m-2n-1}+\lambda_1^2 \plus_{2m+2n+1}\plus_{2m-2n-1}\right)\cdot \mathbb{1} + 2y_1\lambda_1 \minus_{2m}\plus_{2m} \Lambda\ ,\\
\Lambda&:=\begin{pmatrix}
0 & \lambda_1\\
\frac{1}{\lambda_1} & 0
\end{pmatrix}\ .
\end{split}
\end{equation}
It is worthwhile to note the origin of the two elementary matrices
$\mathbb{1}$ and $\Lambda$ in this formula, which is simply the
application of the ``symmetrisation fusion functor'' $Y_{*}$ onto
$y_1\equiv x_1+x_2$ and $\lambda_1\equiv x_1-x_2$ (i.e.\ to $y_1$ upon
embedding into the polynomial ring $\bC[x_1,x_2]$, and to $\lambda_1$
considered as an element of $\bC[x_1,x_2]$):
\begin{equation}
Y_{*} (y_1)=y_1\mathbb{1}\quad ,\quad Y_{*}
(\lambda_1)=\lambda_1\Lambda\ .
\end{equation}
The crucial feature of~\eqref{eq:D0monJn} is the fact that the
off-diagonal entries of $\tilde{D}_{0,m}\big( \cJ_n\big)$ do not
depend on $n$, i.e.\ on the label of the elementary polynomial~$\cJ_n$. 

For later convenience, we will also define the symbol
$\tilde{\tilde{D}}_{0,m}\big(\mathcal{J}_n\big)$ to denote the
following form for $\tilde{D}_{0,m}\big( \mathcal{J}_n \big)$, which
is obtained via a similarity transformation that rescales the
off-diagonal entries\footnote{This similarity
transformation is independent of $n$ and has identical diagonal
blocks, so it satisfies the two criteria specified above}:
\begin{equation}\label{eq:deftildetildeD}
\begin{split}
\tilde{\tilde{D}}_{0,m}\big( \mathcal{J}_n\big)&:=\tilde{\tilde{\cU}}^{(0)}_{m,n}\cdot \left(\tilde{D}_{0,m}\big( \mathcal{J}_n \big)\right)\cdot \tilde{\tilde{\cU}}_{m,n}^{(1)^{-1}}\\
&=
\left(y_1^2\minus_{2m+2n+1}\minus_{2m-2n-1}+\lambda_1^2\plus_{2m+2n+1}\plus_{2m-2n-1}\right)\mathbb{1}\\
&\quad +
\begin{pmatrix}
0 & 4\minus_{2m}^{2}\plus_{2m}^{2}y_{1}\lambda_1^2\\
y_1 & 0
\end{pmatrix}\\
\tilde{\tilde{\cU}}_{m,n}&:=
\cU_{col\times}\left(2;2\minus_{2m}\plus_{2m}\right)\cdot\cU_{row\times}\left(2;\frac{1}{2\minus_{2m}\plus_{2m}}\right)\ .
\end{split}
\end{equation}
We want to take this as our new standard form for $\cJ_{n(1)}$, so we
define
\begin{equation}
\tilde{\cJ}_{n(1)} := \tilde{\tilde{D}}_{0,1}\big(\cJ_{n} \big) \ .
\end{equation}
We can immediately conclude that
\begin{equation}
D_{(1)}\big(\mathcal{J}_{|L,0\rangle} \big) \cong \begin{pmatrix}
\tilde{\mathcal{J}}_{0(1)} & 0 & \cdots &&&&\\
\mathbb{1} & \tilde{\mathcal{J}}_{1(1)} & 0 & \cdots &&&\\
0 & \mathbb{1} & \tilde{\mathcal{J}}_{2(1)} & 0 & \cdots & &\\
\vdots & \ddots & \ddots & \ddots & &&\\
&&&0& \mathbb{1}& \tilde{\mathcal{J}}_{L-1(1)} & 0\\
&&&& 0& \mathbb{1} & \tilde{\mathcal{J}}_{L(1)}
\end{pmatrix}\ .
\end{equation}
Now we have to look for similar expressions for $\cJ_{n(m)}$ for
$m\geq 2$. A tedious computation (some ideas of which are presented in
appendix~\ref{sec:alternativeform}) leads to the following claim: we have found an
alternative form of $Q_{n(m)}$ that we call $\tilde{Q}_{n(m)}$
(related by a similarity transformation) and that satisfies the
following property: denote by $C(p_{1},\dotsc,p_{r};m)$ the cone whose
upper right block is given by
\begin{equation}
C(p_{1},\dotsc,p_{r};m)^{(1)} = \begin{pmatrix}
\tilde{\cJ}_{p_{1}(m)} & 0 & \cdots &&&&\\
\mathbb{1} & \tilde{\cJ}_{p_{2}(m)} & 0 & \cdots &&&\\
0 & \mathbb{1} & \tilde{\cJ}_{p_{3}(m)} & 0 & \cdots & &\\
\vdots & \ddots & \ddots & \ddots & &&\\
&&&0& \mathbb{1}& \tilde{\cJ}_{p_{r-1}(m)} & 0\\
&&&& 0& \mathbb{1} & \tilde{\mathcal{J}}_{p_{r}(m)}
\end{pmatrix} \ ,
\end{equation} 
where as usual $\tilde{\cJ}_{n(m)}$ is the upper right block of
$\tilde{Q}_{n(m)}$. Then 
\begin{equation}\label{fusionofnewstandardform}
D_{(1)}\big( C(p_{1},\dotsc,p_{r};m) \big) \cong 
C(p_{1},\dotsc,p_{r};m-1) \oplus C(p_{1},\dotsc,p_{r};m+1) 
\end{equation}
for generic $p_{1},\dotsc ,p_{r}$. The alternative standard form $\tilde{\cJ}_{n(m)}$ is given by
\begin{subequations}
\label{eq:alternativestandardform}
\begin{align}
&\text{for even $m$:} \nonumber\\
&\ \tilde{\cJ}_{n(m)} ={\begin{pmatrix}
\st\eta^{2m}\mathcal{J}_n & \st 0 & \st \cdots\\
\st \eta^{2m-4}\Psi_{0,2} & \st \eta^{2m-4}\tilde{\tilde{D}}_{0,2}\big(  \mathcal{J}_n\big) & \st 0 & \st \cdots \\
\st 0 & \st \eta^{2m-8}\Psi_{2,4} & \st \eta^{2m-8}\tilde{\tilde{D}}_{0,4}\big( \mathcal{J}_n\big) & \st 0 & \st \cdots \\
\st\vdots & \st\ddots & \st \ddots & \st \ddots & \st \ddots \\
&&&&& \st 0 & \st \Psi_{m-2,m} & \st \tilde{\tilde{D}}_{0,m}\big( \mathcal{J}_n \big)
\end{pmatrix}}\\
&\text{for odd $m$:}\nonumber\\
&\ \tilde{\cJ}_{n(m)} = {\begin{pmatrix}
\st \eta^{2m-2}\tilde{\tilde{D}}_{0,1}\big(  \mathcal{J}_n\big)  & \st 0 & \st \cdots\\
\st \eta^{2m-6}\Psi_{1,3} & \st\eta^{2m-6}\tilde{\tilde{D}}_{0,3}\big( \mathcal{J}_n\big) & \st 0 & \st \cdots \\
\st 0 & \st \eta^{2m-10}\Psi_{3,5} & \st \eta^{2m-10}\tilde{\tilde{D}}_{0,5}\big( \mathcal{J}_n\big) & \st 0 & \st \cdots \\
\st\vdots & \st \ddots & \st \ddots & \st \ddots & \st \ddots \\
&&&&& \st 0 & \st \Psi_{m-2,m} & \st \tilde{\tilde{D}}_{0,m}\big( \mathcal{J}_n\big)
\end{pmatrix}} \ .
\end{align}
\end{subequations}
This formula is obtained by an extrapolation of the pattern one
observes for small values $m$. We expect it to be correct for $m\leq
k+2$ if $k$ is odd, whereas for $k$ even we can from our derivation
only conclude that it should be valid for $m+|p_{i}|\leq k/2$ (see the
discussion in appendix~\ref{sec:finitelevelconstraints}). If on
the other hand our conjecture is correct that the decomposition
in~\eqref{fusionofnewstandardform} can also be done purely by using
similarity transformations that satisfy the two properties formulated
above, i.e.\ by blockwise transformations independent of the label
$p_{i}$, then also the constraint should not depend on the label
$p_{i}$ and we could conclude that the formula is valid for all $m\leq
k/2$.

We can then finally write down a matrix factorisation for a general rational boundary
state $|L,\ell\rangle$ in the form
\begin{equation}\label{eq:QLl}
\cJ_{|L,\ell\rangle} \cong \begin{pmatrix}
\tilde{\mathcal{J}}_{0(\ell)} & 0 & \cdots &&&&\\
\mathbb{1} & \tilde{\mathcal{J}}_{1(\ell)} & 0 & \cdots &&&\\
0 & \mathbb{1} & \tilde{\mathcal{J}}_{2(\ell)} & 0 & \cdots & &\\
\vdots & \ddots & \ddots & \ddots & &&\\
&&&0& \mathbb{1}& \tilde{\mathcal{J}}_{L-1(\ell)} & 0\\
&&&& 0& \mathbb{1} & \tilde{\mathcal{J}}_{L(\ell)}
\end{pmatrix}\ .
\end{equation}
For odd $k$ this formula should hold for all $L$ and $\ell$, whereas
for even $k$ we have constraints. From the discussion above we
conclude that it should be valid at least for $L+\ell\leq k/2$, but if
our conjecture on the similarity transformation is correct, it should
hold for all $\ell \leq k/2$. If this is true then using the
identification
\begin{equation}
|L,\ell \rangle = \overline{|L,k+1-\ell \rangle} 
\end{equation}
one can get a factorisation for every rational boundary state also for
even $k$.

Up to this issue of the constraints due to the level $k$, we have
formulated a \emph{complete dictionary} between
matrix factorisations and rational boundary states for the
Kazama-Suzuki model of type $SU(3)_k/U(2)$.

\subsection{Effects of finite levels}

For a finite level $k$ there are only finitely many rational boundary
states, so that if we continue to apply $D_{(1)}$ we should see
dependencies between the factorisations that arise
due to the identity
\begin{equation}
\eta^{k+3}=-1 \ .
\end{equation} 
Checking the dependencies is then another test that we identified the
correct matrix factorisation.

When we successively determine factorisations by applying the fusion
functor $D_{(1)}$ on factorisations $Q_{|L,\ell\rangle}$ we expect our first
interesting effect for the special value $\ell=\lfloor \frac{k+1}{2}\rfloor$:
\begin{align}
\label{app:evenkspecialfusion}
&k\in2\bZ &D_{(1)}\big( Q_{|L,\tfrac{k}{2}\rangle}\big)&\cong Q_{|L,\tfrac{k}{2}-1\rangle}\oplus
\underset{=\,Q_{\overline{|L,\tfrac{k}{2}\rangle}}}{\underbrace{Q_{|L,\tfrac{k}{2}+1\rangle}}}\\
&k\in2\bZ+1 & D_{(1)}\big( Q_{|L,\tfrac{k+1}{2}\rangle}\big)&\cong Q_{|L,\tfrac{k-1}{2}\rangle}\oplus
\underset{=\,Q_{\overline{|L,\tfrac{k-1}{2}\rangle}}}{\underbrace{Q_{|L,\tfrac{k+3}{2}\rangle}}}\ .
\end{align}
We notice a crucial difference in the cases $k$ odd and $k$ even,
respectively:\footnote{See also Figure~1 of~\cite{Behr:2010ug} for
illustration} for $k$ even, there exists one special irreducible
factor of the superpotential $W^y_{2;k}$ as defined
in~\eqref{eq:WKSdef}, namely the factor
$\mathcal{J}_{\tfrac{k}{2}}=y_1$. We will thus have to discuss the two
cases separately.

For the \emph{case $k$ odd}, all the irreducible factors
$\mathcal{J}_i$ of the superpotential $W_{2;k}^y$ are of the generic
form~\eqref{eq:JpolyDef}, so the only effect of the special label
$\ell=\frac{k+1}{2}$ consists in a number of identifications. For
concreteness, consider the case of the rational matrix factorisations
$Q_{|0,\ell\rangle}$, for which we found earlier the formula
(see~\eqref{eq:rMF})
\begin{equation}\label{app:eq:J0l}
\mathcal{J}_{0({\ell})}=\begin{pmatrix}
\mathcal{J}_{-{\ell}} & 0 & \cdots &&&&&&\\
y_1 & \mathcal{J}_{-{\ell}+2} & 0 & \cdots &&&&&&\\
\chi_{(-{\ell}+2)} & y_1 & \mathcal{J}_{-{\ell}+4} & 0 & \cdots &&&&&\\
0 & \chi_{(-{\ell}+4)} & y_1 & \mathcal{J}_{-{\ell}+6} & 0 & \cdots &&&&\\
\vdots & \ddots & \ddots & \ddots & \ddots & \ddots & &&&\\
\\
&&&&&& 0 & \chi_{({\ell-2})} & y_1 & \mathcal{J}_{{\ell}}
\end{pmatrix}\ ,
\end{equation}
with $\chi_{(p)}$ defined in~\eqref{defchi}. 
Using the obvious identification of labels
\begin{equation}\label{app:eq:Jident}
\mathcal{J}_{-n}=y_1^2\mu_{-2n+1}\mu_{2n-1}+\lambda_1^2\pi_{-2n+1}\pi_{2n-1}=\mathcal{J}_{n-1}\ ,
\end{equation}
we observe that the negative labels in~\eqref{app:eq:J0l} are mapped
to positive labels in such a way that for $\ell=\frac{k+1}{2}$ the
list of diagonal entries of $\mathcal{J}_{|0,\ell\rangle}$ exhausts
the list of all irreducible factors (which are labelled
$\mathcal{J}_0,\mathcal{J}_1,\ldots, \mathcal{J}_{\tfrac{k+1}{2}}$ for
$k$ odd). It may be checked that (unlike in the case of $k$ even,
which will be discussed below) no special relations
play a role when applying $D_{(1)}$ to $Q_{|0,\tfrac{k+1}{2}\rangle}$, i.e.\ we obtain our usual result
\begin{equation}
D_{(1)}\big( Q_{|0,\tfrac{k+1}{2}\rangle}\big)\cong
Q_{|0,\tfrac{k-1}{2}\rangle}\oplus Q_{|0,\tfrac{k+3}{2}\rangle}\ .
\end{equation}
The only structural speciality in $Q_{|0,\tfrac{k+3}{2}\rangle}$ stems from the fact that
\begin{equation}
\mathcal{J}_{-\tfrac{k+3}{2}}=\mathcal{J}_{\tfrac{k+1}{2}}
=\mathcal{J}_{\tfrac{k+3}{2}}\ ,
\end{equation}
which may be checked by inspecting~\eqref{eq:JpolyDef}.
In addition, the relation
\begin{equation}\label{app:eq:chiId}
\chi_{(-m)}=\chi_{(m-1)}\ ,
\end{equation} 
which follows immediately from the definition~\eqref{defchi} of
$\chi_{(m)}$, may be employed to convert every constant $\chi_{(m)}$
with negative label into one with positive label. 
Additional arguments for proving the second part of the claim, i.e.\
that $Q_{|L,\frac{k+3}{2}\rangle}\cong
Q_{\overline{|L,\frac{k-1}{2}\rangle}}$, are introduced below when we discuss the
case of even $k$, but we refrain from carrying out the explicit
computations for brevity, since they are entirely analogous to those
necessary in the more interesting case of $k$ even.  
\smallskip

For the \emph{case $k$ even}, we encounter the problem that the
formula~\eqref{app:eq:J0l} for $\cJ_{0(\ell)}$ is only valid for
$\ell\leq\frac{k}{2}$. Therefore when we want to
check~\eqref{app:evenkspecialfusion}, we cannot directly use the
formula~\eqref{app:eq:J0l} for $Q_{|0,\frac{k}{2}+1\rangle}$. The
problem occurs when $D_{(1)}$ hits the polynomial factor $\cJ_{p}$ with highest $p$ ($p=k/2$) on
the diagonal of $\cJ_{0(k/2)}$. We then have (see~\eqref{eq:D0monJn})
\begin{equation}
\begin{split}\label{app:eq:D1onJnSpecialkEven}
D_{(1)}\big(  \mathcal{J}_{k/2}\big)&=
\left(y_1^2 \mu_{k+3}\mu_{-k+1}+\lambda_1^2 \pi_{k+3}\pi_{-k+1}\right)\cdot \mathbb{1} + y_1\lambda_1 \mu_{4} \Lambda\\
&=\pi_{-2k-2}y_1^2\mathbb{1}+\mu_{-2k-2}y_1\lambda_1\Lambda\ .
\end{split}
\end{equation}
It is now a straightforward computation to demonstrate that via the similarity transformations
\begin{equation}\label{app:defhatU}
\begin{split}
\hat{\cU}_{(k)}&:=
\cU_{col\times}\left(1; \frac{1}{\mu_{-2k-2}}\right)\cdot
\cU_{row\times}\left(1; \mu_{2k+2}\right)\cdot\\
&\quad\cdot\cU_{col}\left(1,2,\frac{y_1 \pi_{2k+2}}{\mu_{2k+2}}\right)\cdot
\cU_{row}\left(1,2,\frac{y_1 \pi_{2k+2}}{\mu_{2k+2}}\right)
\end{split}
\end{equation}
we may realise the isomorphism
\begin{equation}
\begin{split}\label{app:eq:D1onJnSpecialkEvenB}
D_{(1)}\big( \mathcal{J}_{k/2}\big)&\cong
\begin{pmatrix}
0 & y_1 \big(\pi_{2k+2}\pi_{-2k-2}y_1^2+\mu_{2k+2}\mu_{-2k-2}\lambda_1^2\big)\\
y_1 & 0
\end{pmatrix}\\
&=\begin{pmatrix}
0 & y_1 \mathcal{J}_{\frac{k}{2}-1}\\
y_1 & 0
\end{pmatrix}\ .
\end{split}
\end{equation}
Here, we have made use of the fact that
\begin{equation}
\plus_{p} = \minus_{p+k+3} \ .
\end{equation}
We are now in the position to determine
$\cJ_{0(\frac{k}{2}+1)}$ that occurs in the
relation~\eqref{app:evenkspecialfusion} for $L=0$,
\begin{equation}\label{app:eq:SRkEven}
k\in2\bZ:\quad D_{(1)}\circ Q_{|0,\frac{k}{2}\rangle}\cong Q_{|0,\frac{k}{2}-1\rangle}\oplus
Q_{|0,\frac{k}{2}+1\rangle}\overset{!}{\cong}
Q_{|0,\frac{k}{2}-1\rangle}\oplus
Q_{\overline{|0,\frac{k}{2}\rangle}}\ .
\end{equation}
We start from the explicit
formula~\eqref{app:eq:J0l} for the factorisation $Q_{|0,\ell\rangle}$, which reads
using the relations~\eqref{app:eq:Jident}
and~\eqref{app:eq:chiId}: 
\begin{equation}
\mathcal{J}_{0({\ell})} =\begin{pmatrix}
\mathcal{J}_{\ell -1} & 0 & \cdots &&&&&&\\
y_1 & \mathcal{J}_{\ell -3} & 0 & \cdots &&&&&&\\
\chi_{(\ell-3)} & y_1 & \mathcal{J}_{\ell -5} & 0 & \cdots &&&&&\\
0 & \chi_{(\ell-5)} & y_1 & \mathcal{J}_{\ell -7} & 0 & \cdots &&&&\\
\vdots & \ddots & \ddots & \ddots & \ddots & \ddots & &&&\\
\\
&&&&&& 0 & \chi_{(\ell-2)} & y_1 & \mathcal{J}_{\ell}
\end{pmatrix}\ .
\end{equation}
Here, the entries on the diagonal run from $\mathcal{J}_{\ell-1}$ to
$\mathcal{J}_{\ell-m^{*}}$ in steps of two, where 
\begin{equation*}
m^{*}:=\left\{\begin{array}{ll}
\ell & \ell \ \text{odd}\\
\ell -1 & \ell \ \text{even.}
\end{array} \right.
\end{equation*}
Then, if $m^{*}=\ell$, the next diagonal entries after $\mathcal{J}_0$ read
$\mathcal{J}_1,\mathcal{J}_3,\ldots$. Otherwise, we have that
$\mathcal{J}_{\ell-m^{*}}=\mathcal{J}_1$, after which the next entries
read $\mathcal{J}_0,\mathcal{J}_2,\mathcal{J}_4,\ldots$. 

At $k=2$ we obtain
\begin{equation}
\mathcal{J}_{|0,1\rangle}=\begin{pmatrix}
\mathcal{J}_{-1} & 0\\
y_1 & \mathcal{J}_1
\end{pmatrix}\ ,
\end{equation}
and we immediately compute
\begin{equation}
\begin{split}
D_{(1)}\big( \mathcal{J}_{|0,1\rangle} \big)\bigg\vert_{k=2}&=
\begin{SepA}{c|c}
D_{(1)}\big( \mathcal{J}_{{-1}}\big) & 0\\\hline
D_{(1)}[y_1] & D_{(1)}\big( \mathcal{J}_1\big)
\end{SepA}\\
&\overset{\eqref{app:eq:D1onJnSpecialkEven}}{=}
\begin{SepA}{cc|cc}
\left(y_{1}^2 \mu_1\mu_3+\lambda_1^2\pi_1\pi_3\right) & y_1\lambda_1^2\mu_4 & 0 & 0\\
 y_1\mu_4 & \left(y_{1}^2 \mu_1\mu_3+\lambda_1^2\pi_1\pi_3\right) & 0 & 0\\\hline
 y_1 \pi_2 & \lambda_1^2 \mu_2 & y_1^2 \mu_{-1} & y_1\lambda_1^2 \pi_{-1}\\
 \mu_2 & y_1 \pi_2 & y_1 \pi_{-1} & y_1^2\mu_{-1}
\end{SepA}\ .
\end{split}
\end{equation}
Applying the transformation $\hat{\cU}_{k}$ (see~\eqref{app:defhatU})
to the lower right block, and the standard transformation
${}^{1}\cU_{-1}$ (given in~\eqref{cUr}) to the upper left block, we
obtain the intermediate result
\begin{equation}
\begin{split}
D_{(1)}\big( \mathcal{J}_{|0,1\rangle} \big)\bigg\vert_{k=2}&\cong
\begin{SepA}{cc|cc}
\mathcal{J}_1 & 0 & 0 & 0\\
y_1 & \mathcal{J}_0 & 0 & 0\\\hline
y_1 \pi_3 & \mathcal{J}_0 \frac{\mu_1\pi_1}{\mu_{-1}} & 0 & y_1 \mathcal{J}_0\\
2\mu_1\pi_1 & y_1\frac{\pi_1\pi_{-1}}{\pi_3} & y_1 & 0
\end{SepA}\\
&\cong
\begin{SepA}{cc|cc}
\mathcal{J}_1 & 0 & 0 & 0\\
y_1 & \mathcal{J}_0 & 0 & 0\\\hline
0 & \cJ_0 \left(\frac{\mu_1\pi_1}{\mu_{-1}} - \pi_3\right)& 0 & y_1 \mathcal{J}_0\\
2\mu_1\pi_1 & 0 & y_1 & 0
\end{SepA}\\
&\cong
\begin{SepA}{cc|cc}
\mathcal{J}_1 & 0 & 0 & 0\\
y_1 & 0 & 0 & y_1 \mathcal{J}_0\\\hline
0 & \mathcal{J}_0 & 0 & 0\\
1 & 0 & y_1 & 0
\end{SepA}\ .
\end{split}
\end{equation}
It is then immediately obvious that this result can be transformed into the form
\begin{equation}
D_{(1)}\big( \mathcal{J}_{|0,1\rangle} \big)\bigg\vert_{k=2}\cong
\begin{SepA}{cc|cc}
0 & 0 & y_1 \mathcal{J}_1 & 0\\
0 & 0 & -y_1^2 & y_1 \mathcal{J}_0\\ \hline
0 & \mathcal{J}_0 & 0 & 0\\
1 & 0 & 0 & 0
\end{SepA}=\mathcal{J}_{triv}\oplus \mathcal{J}_{|0,0\rangle}\oplus
\mathcal{J}_{\overline{|0,1\rangle}}\ ,
\end{equation}
which provides an explicit check of the
relation~\eqref{app:eq:SRkEven} for $k=2$.

Using the same tools as introduced in the computation of the explicit
formula for the rational factorisations $Q_{|0,\ell\rangle}$, we can compute for
example the next two cases for the level $k$:
\begin{equation}
\begin{split}
D_{(1)}\big(\mathcal{J}_{|0,2\rangle} \big)\bigg\vert_{k=4}&\cong \mathcal{J}_{|0,1\rangle}\oplus 
\begin{SepA}{cc|cc}
\mathcal{J}_{-3} & 0 & 0 & 0\\
y_1 & \mathcal{J}_{-1} & 0 & 0\\
\chi_{(-1)} & y_1 & 0 & y_1 \mathcal{J}_1\\\hline
0 & 1 & y_1 & 0 
\end{SepA}\\
D_{(1)}\big( \mathcal{J}_{|0,3\rangle}\big)\bigg\vert_{k=6}&\cong \mathcal{J}_{|0,2\rangle}\oplus 
\begin{SepA}{ccc|cc}
\mathcal{J}_{-4} & 0 & 0 & 0 & 0\\
y_1 & \mathcal{J}_{-2} & 0 & 0 & 0\\
\chi_{(-2)} & y_1 & \mathcal{J}_0 & 0 & 0\\
0 & \chi_{(0)} & y_1 & 0 & y_1\mathcal{J}_2\\\hline
0 & 0 &  1 & y_1 & 0 
\end{SepA}\ .
\end{split}
\end{equation}
We observe that the largest part of the factorisation
$Q_{|0,\frac{k}{2}+1\rangle}$ is of the form of an ordinary factorisation
$Q_{|0,\ell\rangle}$. From the first three even $k$
examples, we conjecture the formula
\begin{equation}
\mathcal{J}_{|0,\frac{k}{2}+1\rangle}=
\begin{SepA}{cccccccc|cc}
\mathcal{J}_{-\frac{k}{2}-1} & 0 & \cdots &&&&&&&\\
y_1 & \mathcal{J}_{-\frac{k}{2}+1} & 0 & \cdots &&&&&&\\
\chi_{(-\frac{k}{2}+1)} & y_1 & \mathcal{J}_{-\frac{k}{2}+3} & 0 & \cdots &&&&&\\
0 & \chi_{(-\frac{k}{2}+3)} & y_1 & \mathcal{J}_{-\frac{k}{2}+5} & 0 & \cdots &&&&\\
\vdots & \ddots & \ddots & \ddots & \ddots & \ddots & &&&\\
&&&&&&&&&\\
&&&&&& y_1 & \mathcal{J}_{\frac{k}{2}-3} & 0 & 0\\
&&&&&& \chi_{(\frac{k}{2}-3)} & y_1 & 0 & y_1 \mathcal{J}_{\frac{k}{2}-1}\\[1mm]\hline
&&&&&& 0 & 1 & y_1 & 0
\end{SepA}\ .
\end{equation} 
It remains to check that
\begin{equation}\label{check}
Q_{|0,\frac{k}{2}+1\rangle}\cong
Q_{\overline{|0,\frac{k}{2}\rangle}}\ ,
\end{equation}
or in other words that the upper right block $\cJ_{|0,\frac{k}{2}+1\rangle}$ of
$Q_{|0,\frac{k}{2}+1\rangle}$ can be transformed to the lower left
block $\cE_{|0,\frac{k}{2}\rangle}$ of the factorisation
$Q_{|0,\frac{k}{2}+1\rangle}$ by elementary row and column
operations. 

Let us consider the example $k=4$.  Note that
$\cJ_{|0,3\rangle}$ has two constant entries that we can use to remove
all other entries in their rows and columns, and we obtain 
\begin{equation}
\begin{split}
\mathcal{J}_{|0,3\rangle}\bigg\vert_{k=4}&= 
\begin{SepA}{cc|cc}
\mathcal{J}_{2} & 0 & 0 & 0\\
y_1 & \mathcal{J}_{0} & 0 & 0\\
\chi_{(0)} & y_1 & 0 & y_1 \mathcal{J}_1\\\hline
0 & 1 & y_1 & 0 
\end{SepA}\\
&\cong
\begin{SepA}{cc|cc}
 0 & 0 & -y_1^2 \mathcal{J}_{2} & y_1 \mathcal{J}_{2}\mathcal{J}_1\\
 0 & 0   & y_1 \left(y_1^2-\mathcal{J}_{0}\chi_{(0)} \right)& -y_1^2\mathcal{J}_1\\ \hline
0 & 1 & 0 & 0\\
1 & 0 & 0 & 0 
\end{SepA}\ .
\end{split}
\end{equation}
The upper right block of this last form coincides with
$\widehat{\mathcal{E}}_{0(2)}$ given in~\eqref{app:hatE}, which
therefore proves the relation~\eqref{check} in this case. Similarly we
have verified~\eqref{check} explicitly also for $k=6$. We take this
as another convincing check that we identified the correct matrix
factorisations.

\section{Conclusion and outlook}

In this article we have constructed matrix factorisations for rational
boundary conditions in the $SU(3)/U(2)$ Kazama-Suzuki models. For the
construction it was essential to identify the rational defect
$D_{[(0,0),0;1,3]}$ in the Landau-Ginzburg description. Fusing this
defect to boundary conditions $|L,0\rangle$, one can generate all
boundary conditions $|L,\ell\rangle$. Therefore by fusing the defect
in the Landau-Ginzburg description to the matrix factorisations
describing $|L,0\rangle$, we can obtain all others.

To actually construct these matrix factorisations, it is important to
have an efficient way of computing the fusion. We found an
operator-like description for the fusion of the defect factorisation
corresponding to $D_{[(0,0),0;,1,3]}$ to another factorisation (see~\eqref{defofD}), which
is given by a specific operation on each entry of the
factorisation. In this way we worked out the \emph{matrix factorisations for
all rational boundary conditions} $|L,\ell\rangle$, and hence have obtained a conjecture for a \emph{complete dictionary} between the Landau-Ginzburg formulation and the rational conformal field theory description of the $SU(3)/U(2)$ Kazama-Suzuki models. More precisely, we proved our formula~\eqref{eq:rMF} (and the alternative compact version~\eqref{twotworealisation}) for the matrix factorisations $Q_{|0,\ell\rangle}$ explicitly, while for the $Q_{|L,\ell\rangle}$ factorisations with $L>0$ we have extrapolated the pattern we have observed for small values of $L$ to derive the conjecture for their explicit form (see~\eqref{eq:QLl}). Additional support for our conjecture comes from a detailed discussion of the effects of finite levels $k$, which are consistent with the expectations from the conformal field theory side of the dictionary. We will report in~\cite{NBSFtwo2014} a number of further structural arguments in favor of our conjecture.

Operator-like defects turn out to be very important for explicit
computations. The process of fusing a defect factorisation of
$W(x)-W(\tilde{x})$ to some matrix factorisation of $W(\tilde{x})$ is
described by the tensor product, resulting in a factorisation of
$W(x)$. This tensor product still contains the variables
$\tilde{x}$. To eliminate these auxiliary variables can be a
complicated task, though there are some strategies and algorithms
known how this can be
done~\cite{Brunner:2007qu,Carqueville:2011zea}. For operator-like
defects such as $D_{(1)}$, this step does not have to be
performed -- the process of fusing it to another factorisation is
implemented by a functor that acts on the category of modules over a
polynomial ring.\footnote{Of course, tensoring a defect matrix
factorisation $D$ always defines a functor in the category of matrix
factorisations. The functors we are considering, however, act on the
category of ring modules, and their action on a matrix factorisation
$Q$ is simply given by applying $D$ on $Q$ seen as a ring module
homomorphism.} In this functorial language one can also realise the
morphisms of operator-like defects as morphisms between the
corresponding functors, and in this way one can even define cones of
functors in certain situations. This will be presented in~\cite{NBSFthree2014}.

For the $SU(3)/U(2)$ Kazama-Suzuki models it turns out that all
rational B-type defects can be realised as operator-like defects with
corresponding fusion functors~\cite{NBSFtwo2014}. This
then opens the possibility to study the fusion semi-ring of these defects.
The fusion of rational defects is given by the rational
fusion rules, and with the functorial description one can then
identify the rational semi-ring structure also in the
Landau-Ginzburg description. We will report on this in an upcoming
publication~\cite{NBSFtwo2014}.
\smallskip

After having the $SU(3)/U(2)$ model under control, one may ask whether
a similar strategy also works for the higher rank models. Also in this
case there exists a variable transformation interface to a product of minimal
models~\cite{Behr:2012xg}, and the natural ansatz would be to study the effect
of fusing it to known factorisations in the minimal models, maybe to
the permutation factorisations of~\cite{Enger:2005jk}. Although it is
far from obvious, one might be lucky and generate in this way
factorisations for rational boundary conditions or defects.

\section*{Acknowledgements}

We thank Nils Carqueville, Dan Murfet and Ingo Runkel for helpful discussions.

\appendix

\section{Similarity transformations}\label{sec:Utrafos}

For convenience we summarise here our conventions for the basic row
and column transformations. For a matrix factorisation $Q$ of matrix
size $2d$ of the form
\begin{equation}
Q=\begin{pmatrix}
0 & Q^{(1)}\\
Q^{(0)} & 0
\end{pmatrix}
\end{equation}
with the two $d\times d$ blocks $Q^{(0)}$ and $Q^{(1)}$, similarity
transformations are given by invertible $2d\times 2d$ matrices $\cU$
of the form
\begin{equation}
\cU =\begin{pmatrix}
\cU^{(0)} & 0\\
0 & \cU^{(1)}
\end{pmatrix}\ .
\end{equation} 
They act on $Q$ as
\begin{equation}
Q \mapsto \cU \cdot Q \cdot \cU^{-1} = 
\begin{pmatrix}
0 & \cU^{(0)}\cdot Q^{(1)} \cdot (\cU^{(1)})^{-1}\\
\cU^{(1)}\cdot Q^{(0)} \cdot (\cU^{(0)})^{-1} & 0
\end{pmatrix}\ .
\end{equation}
The group of similarity transformations can be generated by elementary row
and column transformations on $Q^{(1)}$ (which induce corresponding elementary
column and row transformations on $Q^{(0)}$). For the basic operations
we take
\begin{subequations}\label{eq:Ulist}
\begin{align}
\cU_{row}(r,s;p)_{ij} &:= \delta_{i,j} + p\,\delta_{i,s}\delta_{j,r} \quad
(p\ \text{any polynomial})\\
&\text{adds row $r$ multiplied by $p$ to row $s$ in $Q^{(1)}$}\nonumber\\
&\text{adds column $s$ multiplied by $-p$ to column $r$ in
$Q^{(0)}$}\nonumber\\[2mm]
\cU_{row\times} (r;\alpha)_{ij} &:= \delta_{ij} (1 +\delta_{i,r} (\alpha -1))
\quad (\alpha \in \mathbb{C})\\
&\text{multiplies row $r$ with $\alpha$ in $Q^{(1)}$}\nonumber\\
&\text{multiplies column $r$ with $1/\alpha$ in $Q^{(0)}$}\nonumber\\[2mm]
\cU_{col} (r,s;p)_{ij} &:= \delta_{i,j} - p\,\delta_{i,r+d}\delta_{j,s+d}\quad
(p\ \text{any polynomial})\\
&\text{adds column $r$ multiplied by $p$ to column $s$ in $Q^{(1)}$}\nonumber\\
&\text{adds row $s$ multiplied by $-p$ to row $r$ in
$Q^{(0)}$}\nonumber\\[2mm]
\cU_{col\times} (r;\alpha)_{ij} &:= \delta_{ij} \left( 1+\delta_{i,d+r}
\left( \frac{1}{\alpha} -1)\right)\right)
\quad (\alpha \in \mathbb{C})\\
&\text{multiplies column $r$ with $\alpha$ in $Q^{(1)}$}\nonumber\\
&\text{multiplies row $r$ with $1/\alpha$ in $Q^{(0)}$.}\nonumber
\end{align}
\end{subequations}

\section{Multiple defect action on polynomial factorisations}\label{app:B}

In this appendix we want to prove that the factorisations $Q_{n(m)}$
with upper right block
\begin{equation}\label{app:defQnm}
\begin{split}
Q^{(1)}_{n(m)} =\cJ_{n(m)}&=\begin{pmatrix}
\cJ_{n-m} & 0 & \cdots &&&&&&\\
y_1 & \cJ_{n-m+2} & 0 & \cdots &&&&&&\\
\chi_{(n-m+2)} & y_1 & \cJ_{n-m+4} & 0 & \cdots &&&&&\\
0 & \chi_{(n-m+4)} & y_1 & \cJ_{n-m+6} & 0 & \cdots &&&&\\
\vdots & \ddots & \ddots & \ddots & \ddots & \ddots & &&&\\
\\
&&&&&& 0 & \chi_{(n+m-2)} & y_1 & \cJ_{n+m}
\end{pmatrix}
\end{split}
\end{equation}
have the following behaviour when we apply the fusion functor
$D_{(1)}$:
\begin{equation}
D_{(1)} Q_{n(m)} \cong  Q_{n(m-1)}\oplus Q_{n(m+1)}\ ,
\end{equation}
where it is understood that $Q_{n(-1)}$ is omitted for $m=0$. We have
proven this relation for $m=0$ and $m=1$ already in the main
text. From the form~\eqref{app:defQnm} we see that the
factorisations $Q_{n(m)}$ contain the factorisations $Q_{n-m},\dotsc ,Q_{n+m}$ as
building blocks. For $Q_{n}$ we have shown (see~\eqref{Qn1}) that
\begin{equation}
D_{(1)} (Q_{n(0)}) = \cU_{n}^{-1} \cdot Q_{n(1)} \cdot \cU_{n} \ .
\end{equation}
When we apply $D_{(1)}$ on $Q_{n(m)}$, we will apply the similarity
transformations $\cU_{j}$ on each block $D_{(1)}(Q_{j})$ that
appears. We find
\begin{align}
&\cU_{n(m)}^{a(0)} \cdot D_{(1)} (\cJ_{n(m)}) \cdot (\cU_{n(m)}^{a(1)})^{-1} \nonumber\\
&=  \begin{pmatrix}
\cJ_{n-m(1)} & 0 & \cdots &&&&&&\\
\Phi^{A}_{n-m+1} & \cJ_{n-m+2(1)} & 0 & \cdots &&&&&&\\
\Phi^{B}_{n-m+2} & \Phi^{A}_{n-m+3} & \cJ_{n-m+4(1)} & 0 & \cdots &&&&&\\
0 & \Phi^{B}_{n-m+4} & \Phi^{A}_{n-m+5} & \cJ_{n-m+6(1)} & 0 & \cdots &&&&\\
\vdots & \ddots & \ddots & \ddots & \ddots & \ddots & &&&\\
\\
&&&&&& 0 & \Phi^{B}_{n+m-2} & \Phi^{A}_{n+m-1} & \cJ_{n+m(1)}
\end{pmatrix}  
\label{app:D1cJnm}
\end{align}
with the similarity transformation $\cU^{a}_{n(m)}$ given by
\begin{equation}
\cU^{a}_{n(m)} = \prod_{j=0}^{m}{}^{1+2j}\cU_{n-m+2j} \ . 
\end{equation}
Here the left upper index on the $\cU_{n}$ indicates on which row or
column the transformation acts (see~\eqref{cUr}). The blocks
$\Phi^{A/B}_{j}$ are given by
\begin{equation}
\Phi^{A}_{j} = \cU_{j+1}^{(0)} \cdot D_{(1)}(y_{1}) \cdot
(\cU_{j-1}^{(1)})^{-1} = \begin{pmatrix}
y_{1}\frac{\plus_{2j+3}}{\plus_{2j+5}} & \cJ_{j}\frac{\minus_{2}\,\minus_{4}}{\plus_{2j+5}\,\plus_{-2j+3}}\\
\frac{1}{2\plus_{2}} & y_{1}\frac{\plus_{-2j+1}}{\plus_{-2j+3}}
\end{pmatrix}\ ,
\end{equation}
and
\begin{equation}
\Phi^{B}_{j} = \cU_{j+2}^{(0)} \cdot D_{(1)} (\chi_{(j)}) \cdot (\cU_{j-2}^{(1)})^{-1}= \begin{pmatrix}
\frac{\plus_{2j+5}\,\plus_{-2j+3}}{4\plus_{2}^{2}\,\plus_{-2j+1}\,\plus_{2j+7}} &
-y_{1}\frac{\minus_{4}\,\minus_{-4}\,\plus_{2j+5}\,\plus_{-2j+3}}{4\plus_{2}^{2}\,
\plus_{2j+7}\,\plus_{-2j+5}\,\plus_{2j-1}\,\plus_{-2j-3}}
\\
0 & \frac{\plus_{2j+5}\,\plus_{-2j+3}}{4\plus_{2}^{2}\,\plus_{2j+3}\,\plus_{-2j+5}}
\end{pmatrix}\ .
\end{equation}
We now reorganise the result~\eqref{app:D1cJnm} into the block form
\begin{align}
&\cU_{n(m)}^{a(0)} \cdot D_{(1)} (\cJ_{n(m)}) \cdot (\cU_{n(m)}^{a(1)})^{-1} \nonumber\\
&=
\left(\!\begin{array}{cccccccccccc}
\cJ_{{n-m-1}} & 0 & \cdots &  & & & &  &  &  &  & \\
\Psi^{A,a} & M^{a}_{n-m+1} & 0 & \cdots & & &  &  &  &  &  & \\
\Psi^{B,a}_{n-m+1} & \Psi^{C,a}_{n-m+2} & M^{a}_{n-m+3} & 0 & \cdots & &  &  &  &  &  & \\
0 & \Psi^{D,a}_{n-m+3} & \Psi^{C,a}_{n-m+4} & M^{a}_{n-m+5} & 0 & & &  &  &  &  & \\
\vdots & \ddots & \ddots & \ddots & \ddots &  &  &  & & & &\\
\\
& &  &  &  &  &  & 0 & \Psi^{D,a}_{n+m-3} & \Psi^{C,a}_{n+m-2} & M^{a}_{n+m-1} & 0 \\
& &  &  &  &  & & & 0 &\Psi^{B',a}_{n+m-1} & \Psi^{A',a} & \cJ_{{n+m+1}}
\end{array}\right)
\end{align}
with
\begin{subequations}
\begin{align}
M_{p}^{a} &= \begin{pmatrix}
\cJ_{p} & 0 \\
\frac{\minus_{2}\,\minus_{4}}{\plus_{2p+5}\,\plus_{-2p+3}}\cJ_{p} & \cJ_{p}
\end{pmatrix} & &\begin{array}{ll}
\text{row:} & p-n+m+1\\
\text{column:} & p-n+m+1
\end{array}\\
& (p=n-m-1+2r,\ r=1,\dotsc ,m)\nonumber\\[2mm]
\Psi^{A,a} &= y_{1}\begin{pmatrix}
1 \\
\frac{\plus_{2n-2m+5}}{\plus_{2n-2m+7}}
\end{pmatrix} & & \begin{array}{ll}
\text{row:} & 2\\
\text{column:} & 1
\end{array}\\[2mm]
\Psi^{A',a} &= y_{1}\begin{pmatrix}
\frac{\plus_{2n-2m+5}}{\plus_{2n-2m+7}} & 1
\end{pmatrix} && \begin{array}{ll}
\text{row:} & 2m+2\\
\text{column:} & 2m
\end{array}\\[2mm]
\Psi^{B,a}_{n-m+1} &= \begin{pmatrix}
\frac{1}{2\plus_{2}}\\
\frac{\plus_{2n-2m+9}\,\plus_{-2n+2m-1}}{4\plus_{2}^{2}\,\plus_{-2n+2m-3}\,\plus_{2n-2m+11}}
\end{pmatrix} & & \begin{array}{ll}
\text{row:} & 4\\
\text{column:} & 1 
\end{array}\\[2mm]
\Psi^{B',a}_{n+m-1} &= \begin{pmatrix}
\frac{\plus_{2n+2m+1}\,\plus_{-2n-2m+7}}{4\plus_{2}^{2}\,\plus_{2n+2m-1}\,\plus_{-2n-2m+9}}
& \frac{1}{2\plus_{2}}
\end{pmatrix} && \begin{array}{ll}
\text{row:} & 2m+2\\
\text{column:} & 2m-2
\end{array}\\[2mm]
\Psi^{C,a}_{p} &= y_{1}\begin{pmatrix}
 \frac{\plus_{-2p+3}}{\plus_{-2p+5}}& 1\\
 -\frac{\minus_{4}\,\minus_{-4}\,\plus_{2p+5}\,\plus_{-2p+3}}{4\plus_{2}^{2}
\plus_{2p+7}\,\plus_{-2p+5}\,\plus_{2p-1}\,\plus_{-2p-3}}& \frac{\plus_{2p+5}}{\plus_{2p+7}} 
\end{pmatrix} && \begin{array}{ll}
\text{row:} & p-n+m+2\\
\text{column:} & p-n+m
\end{array}\\
&(p=n-m+2r,\ r=1,\dotsc ,m-1)\nonumber\\[2mm]
\Psi^{D,a}_{p}&= \begin{pmatrix}
 \frac{\plus_{2p+3}\,\plus_{-2p+5}}{4\plus_{2}^{2}\,\plus_{2p+1}\,\plus_{-2p+7}}& \frac{1}{2\plus_{2}}\\
0 & \frac{\plus_{2p+7}\,\plus_{-2p+1}}{4\plus_{2}^{2}\,\plus_{-2p-1}\,\plus_{2p+9}}
\end{pmatrix} && \begin{array}{ll}
\text{row:} & p-n+m+3\\
\text{column:} & p-n+m-1 \ .\end{array}\\
&(p=n-m+2r+1,\ r=1,\dotsc ,m-2) \nonumber
\end{align}
\end{subequations}
Here we always stated the row and column number of the upper left
entry of the given block.

Our strategy is now to eliminate all diagonal terms in the blocks
$M,\Psi^{C}$ and $\Psi^{D}$ by similarity transformations (and the
bottom/left entries in $\psi^{A},\Psi^{B}/\Psi^{A'},\Psi^{B'}$). If we
can achieve this, the factorisation will split into a direct sum of
two factorisations.

We start with the blocks $\Psi^{C}$. The similarity transformations
\begin{equation}
\cU^{b}_{n(m)}= \prod_{j=1}^{m} {}^{2j}\cU_{n-m+2j-1(1)} 
\end{equation}
(see~\eqref{defcUb}) eliminate their diagonal entries,
\begin{align}
\Psi^{C,b}_{p} &= {}^{p-n+m+2}\cU_{p+1(1)}^{(0)}\cdot \Psi^{C,a}_{p}\cdot \big( {}^{p-n+m}\cU_{p-1}^{(1)}\big)^{-1}\\
&=
\cU_{row}^{(0)}\left(1,2;-\frac{\plus_{2p+5}}{\plus_{2p+7}}\right)
\cdot y_{1}\begin{pmatrix}
 \frac{\plus_{-2p+3}}{\plus_{-2p+5}}& 1\\
 -\frac{\minus_{4}\,\minus_{-4}\,\plus_{2p+5}\,\plus_{-2p+3}}{4\plus_{2}^{2}\,
\plus_{2p+7}\,\plus_{-2p+5}\,\plus_{2p-1}\,\plus_{-2p-3}}& \frac{\plus_{2p+5}}{\plus_{2p+7}} 
\end{pmatrix} \nonumber\\
&\mspace{350mu} \cdot
\left(\cU_{col}^{(1)} \left(2,1;-\frac{\plus_{-2p+3}}{\plus_{-2p+5}} \right)\right)^{-1}\\
&= y_{1} \begin{pmatrix}
0 & 1\\
\kappa_{p} & 0
\end{pmatrix} \ ,
\end{align}
with 
\begin{align}
\kappa_{p}&:= \frac{\delta}{4\plus_{2}^{2}\,\chi_{(p-1)}\chi_{(p+1)}}\\
\delta &:= -\frac{1}{4\plus_{2}\plus_{-2}} =
-\frac{\eta^{2}}{(1+\eta^{2})^{2}} \ .
\end{align}
Also one of the entries in $\Psi^{A}$ and $\Psi^{A'}$ is eliminated,
\begin{align}
\Psi^{A,b} &= \begin{pmatrix}
y_{1} \\
0
\end{pmatrix} & \Psi^{A',b} &= \begin{pmatrix}
0 & y_{1}
\end{pmatrix} \ .
\end{align}
The effect on the other blocks is
\begin{subequations}
\begin{align}
M_{p}^{b} &=  \begin{pmatrix}
\cJ_{p} & 0 \\
-\frac{1}{2\plus_{2}\chi_{(p)}}\cJ_{p} & \cJ_{p}\end{pmatrix}\\
\Psi^{B,b}_{n-m+1} &= \begin{pmatrix}
\frac{1}{2\plus_{2}}\\
\kappa_{n-m+2}\chi_{(n-m+1)}
\end{pmatrix} & \Psi^{B',b}_{n+m-1} &= \begin{pmatrix}
\kappa_{n+m-2} \chi_{(n+m-1)}
& \frac{1}{2\plus_{2}}
\end{pmatrix}\\
\Psi^{D,b}_{p}&= \begin{pmatrix}
\kappa_{p-1}\chi_{(p)}& \frac{1}{2\plus_{2}}\\
0 & \kappa_{p+1}\chi_{(p)}
\end{pmatrix}\ .
\end{align}
\end{subequations}
We then turn to the blocks $M$ and apply the transformation
\begin{equation}
\cU^{c}_{n(m)} = \prod_{j=1}^{m} {}^{2j}_{\ c}\cU_{n-m+2j-1(1)}
\end{equation}
(see~\eqref{defcUc}), whose effect on the blocks $M_{p}$ is
\begin{align}
M_{p}^{c} &= \cU_{row}^{(0)}\left( 2,1;2\plus_{2}\chi_{(p)}\right) \cdot
 \begin{pmatrix}
\cJ_{p} & 0 \\
-\frac{1}{2\plus_{2}\chi_{(p)}}\cJ_{p} & \cJ_{p}\end{pmatrix}
\cdot \left(\cU_{col}^{(1)}\left(1,2; 2\plus_{2}\chi_{(p)} \right)\right)^{-1}\\
&= \begin{pmatrix}
0 & 2\plus_{2}\chi_{(p)}\cJ_{p} \\
-\frac{1}{2\plus_{2}\chi_{(p)}}\cJ_{p} & 0 \end{pmatrix} \ .
\end{align}
We then rescale the entries by a further similarity transformation
given by
\begin{equation}\label{app:defUd}
\begin{split}
\cU^{d}_{n(m)} = & \prod_{j=1}^{m} \Bigg\{
\cU_{row\times}\left(2j;\prod_{l=0}^{m-j}\frac{1}{2\plus_{2}\chi_{(n+m-2l-1)}}
\right) 
\cU_{row\times}\left(2j-1;\prod_{l=0}^{m-j}\frac{1}{2\plus_{2}\chi_{(n+m-2l-1)}}
\right) \\
&\quad\cdot
\cU_{col\times}\left(2j;\prod_{l=0}^{m-j}2\plus_{2}\chi_{(n+m-2l-1)}\right)
\cU_{col\times}\left(2j-1;\prod_{l=0}^{m-j}2\plus_{2}\chi_{(n+m-2l-1)}\right)
\Bigg\} \ ,
\end{split}
\end{equation}
and the blocks read after this transformation
\begin{subequations}
\begin{align}
M^{d}_{p} &= \begin{pmatrix}
0 & \cJ_{p}\\
-\cJ_{p} & 0
\end{pmatrix} \\
\Psi^{A,d} &= \begin{pmatrix}
y_{1} \\
0
\end{pmatrix} & \Psi^{A',d} &= \begin{pmatrix}
0 & y_{1}
\end{pmatrix}\\
\Psi^{B,d}_{n-m+1} &= \chi_{(n-m+1)}\begin{pmatrix}
1+\delta\\
\delta
\end{pmatrix} & \Psi^{B',d}_{n+m-1} &= \chi_{(n+m-1)}\begin{pmatrix}
\delta 
& 1+\delta
\end{pmatrix}\\
\Psi^{C,d}_{p} &=y_{1} \begin{pmatrix}
\delta  & 1+\delta \\
\delta  & \delta 
\end{pmatrix} \\
\Psi^{D,d}_{p}&= \chi_{(p)}\begin{pmatrix}
\delta & 1+2\delta \\
0 & \delta 
\end{pmatrix}\ .
\end{align}
\end{subequations}
For the next step we introduce another symbol, $\Delta_{p}$, that we
define recursively by
\begin{equation}\label{app:defofDelta}
\Delta_{p+1} = 1+\frac{\delta}{\Delta_{p}} \quad ,\quad \Delta_{1}=1\ .
\end{equation}
We now want to eliminate the lower right entries of $\Psi^{C}$ and
$\Psi^{D}$, and we perform the similarity transformations
\begin{align}\label{app:defUe}
\cU^{e}_{n(m)}&:=\prod_{r=1}^{m-1}   \cU_{rc}\left(2r+2,2r+3;-\frac{\delta}{\Delta_{r}\Delta_{r+1}}\right)\\
\cU_{rc}\left(r,s;p\right)&:=\cU_{col}\left(r,s;p\right)\cdot
\cU_{row}\left(r,s;p\right) \ .
\end{align}
Because we do the same transformation on the rows and on the columns,
the blocks $M_{p}^{d}$ will be left unchanged, $M_{p}^{e}=M_{p}^{d}$. 
The other blocks transform to
\begin{subequations}
\begin{align}
\Psi^{A,e} &= \begin{pmatrix}
y_{1} \\
0
\end{pmatrix} & \Psi^{A',e} &= \begin{pmatrix}
0 & y_{1}
\end{pmatrix}\\
\Psi^{B,e}_{n-m+1} &= \chi_{(n-m+1)}\begin{pmatrix}
1+\delta\\
0
\end{pmatrix} & \Psi^{B',e}_{n+m-1} &= \chi_{(n+m-1)}\begin{pmatrix}
\delta 
& \Delta_{m}
\end{pmatrix}\\
\Psi^{C,e}_{n-m+2r}  &=y_{1} \begin{pmatrix}
\delta  & \Delta_{r+1} \\
\frac{\delta}{\Delta_{r+1}}  & 0 
\end{pmatrix} \\
\Psi^{D,e}_{n-m+2r+1}&= \chi_{(n-m+2r+1)}\begin{pmatrix}
\delta   & \Delta_{r+1}\Delta_{r+2}  \\
-\frac{\delta^{2}}{\Delta_{r+1}\Delta_{r+2}}  & 0
\end{pmatrix}\ .\mspace{-200mu}
\end{align}
\end{subequations}
To formulate our final similarity transformation we introduce the
quantity $\gamma_{p,q}$ defined as
\begin{equation}
\gamma_{p,q}:=
\left(\prod_{i=1}^{q}\frac{\Delta_{p+i}}{\Delta_{i}}\right) \ .
\end{equation}
It has the properties
\begin{subequations}
\begin{align}
\gamma_{p,q} &= \gamma_{q,p} \\
\gamma_{0,q} &=\gamma_{p,0} = 1\\
\gamma_{1,q} &= \Delta_{q+1}\\
\gamma_{p,q-1} &= \frac{\Delta_{q}}{\Delta_{p}}\gamma_{p-1,q} \\
\gamma_{p,q} &= 1+\frac{\delta}{\Delta_{p}\Delta_{q}} \quad (p,q\geq
1)\ . \label{app:gammaproperty5}
\end{align}
\end{subequations}
Whereas the first properties are obvious from the definition of
$\gamma_{p,q}$, we present the proof of the last one:\\
We prove~\eqref{app:gammaproperty5} by induction. We first note that
it is satisfied for $p=1$,
\begin{equation}
\gamma_{1,q} = \Delta_{q+1} = 1+\frac{\delta}{\Delta_{q}} =
1+\frac{\delta}{\Delta_{1}\Delta_{q}}\ ,
\end{equation}
where we used the recursive definition of $\Delta_{q+1}$
(see~\eqref{app:defofDelta}). Now assume
that~\eqref{app:gammaproperty5} holds for some $p\geq 1$. Then
\begin{align}
\gamma_{p+1,q} &= \frac{\Delta_{q+1}}{\Delta_{p+1}} \gamma_{p,q+1}\\
 &= \frac{\Delta_{q+1}}{\Delta_{p+1}}
 \left(1+\frac{\delta}{\Delta_{p}\Delta_{q+1}} \right)\\
 &= \frac{1}{\Delta_{p+1}}\left(\Delta_{q+1}+\frac{\delta}{\Delta_{p}} \right)\\
 &= \frac{1}{\Delta_{p+1}}\left(\frac{\delta}{\Delta_{q}} + \Delta_{p+1} \right)\\
&= 1+\frac{\delta}{\Delta_{p+1}\Delta_{q}}\ .\qquad \qquad \blacksquare
\end{align}
We can finally formulate the transformation that will remove the
remaining diagonal entry in $\Psi^{C}$ and $\Psi^{D}$, which is given by
\begin{equation}\label{app:defUf}
\cU^{f}_{n(m)} := \prod_{r=1}^{m-1} \cU_{rc}
\left(2r+1,2r;-\frac{\delta}{\Delta_{r+1}\gamma_{r+1,m-r-1}} \right) \ .
\end{equation}
We obtain
\begin{subequations}
\begin{align}
M^{f}_{p} &= \begin{pmatrix}
0 & \cJ_{p}\\
-\cJ_{p} & 0
\end{pmatrix} \\
\Psi^{A,f} &= \begin{pmatrix}
y_{1} \\
0
\end{pmatrix} & \Psi^{A',f} &= \begin{pmatrix}
0 & y_{1}
\end{pmatrix}\\
\Psi^{B,f}_{n-m+1} &= \chi_{(n-m+1)}\begin{pmatrix}
1+\delta\\
0
\end{pmatrix} & \Psi^{B',f}_{n+m-1} &= \chi_{(n+m-1)}\begin{pmatrix}
0
& \Delta_{m}
\end{pmatrix}\\
\Psi^{C,f}_{n-m+2r}  &=y_{1} \begin{pmatrix}
0  & \Delta_{r+1} \\
\frac{\delta}{\Delta_{r+1}}  & 0 
\end{pmatrix} \\
\Psi^{D,f}_{n-m+2r+1}&= \chi_{(n-m+2r+1)}\begin{pmatrix}
0   & \Delta_{r+1}\Delta_{r+2}  \\
-\frac{\delta^{2}}{\Delta_{r+1}\Delta_{r+2}}  & 0
\end{pmatrix}\ .\mspace{-200mu} & & 
\end{align}
\end{subequations}
We have thus achieved our goal to eliminate all diagonal terms in
the blocks, and the factorisation can now be decomposed into
two (see table~\ref{table:decomposition} on
page~\pageref{table:decomposition}). Up to remaining multiplicative
transformations of rows and columns, these two factorisations are
precisely $Q_{n(m-1)}$ and $Q_{n(m+1)}$. This proves our claim. \hfill $\blacksquare$ 
\begin{table}[p]
\centering
\begin{sideways}%
\parbox{20cm}{%
{\footnotesize
\begin{align*}
&D_{(1)} \left(\cJ_{n(m)} \right) \cong \\[5mm]
&\left( \begin{array}{cccccccc}
{\color{first} \cJ_{n-m-1}} & {\color{rest}0} & {\color{first}0} & {\color{rest}0} & {\color{first}0} & {\color{rest}0} & {\color{first}0} & \cdots \\
{\color{first} y_{1}} & {\color{rest}0} & {\color{first}\cJ_{n-m+1}} & {\color{rest}0} & {\color{first}0} & {\color{rest}0} & {\color{first}0} & \cdots \\
{\color{rest}0} & {\color{second} -\cJ_{n-m+1}} & {\color{rest}0} & {\color{second}0} & {\color{rest}0} & {\color{second}0} & {\color{rest}0} & \cdots \\
 {\color{first}\chi_{(n-m+1)} (1+\delta)} & {\color{rest}0} & {\color{first} y_{1}\Delta_{2}} & {\color{rest}0} & {\color{first}\cJ_{n-m+3}} & {\color{rest}0} & {\color{first}0} & \cdots \\
{\color{rest}0} & {\color{second}y_{1}\frac{\delta}{\Delta_{2}}} & {\color{rest}0} & {\color{second} -\cJ_{n-m+3}} & {\color{rest}0} & {\color{second}0} & {\color{rest}0} & \cdots \\
{\color{first}0} & {\color{rest}0} & {\color{first}\chi_{(n-m+3)} \Delta_{2}\Delta_{3}} & {\color{rest}0} & {\color{first}y_{1}\Delta_{3}} & {\color{rest}0} & {\color{first}\cJ_{n-m+5}} & \cdots  \\
{\color{rest}0} & {\color{second} -\chi_{(n-m+3)}\frac{\delta^{2}}{\Delta_{2}\Delta_{3}}} & {\color{rest}0} & {\color{second} y_{1}\frac{\delta}{\Delta_{3}}} & {\color{rest}0} & {\color{second} -\cJ_{n-m+5}} & {\color{rest}0} & \cdots \\
{\color{first}0} & {\color{rest}0} & {\color{first}0} & {\color{rest}0} & {\color{first}\chi_{(n-m+5)}\Delta_{3}\Delta_{4}} & {\color{rest}0} & {\color{first}y_{1}\Delta_{4}} & \cdots \\
{\color{rest}0} &{\color{second}0}  & {\color{rest}0} & {\color{second} -\chi_{(n-m+5)}\frac{\delta^{2}}{\Delta_{3}\Delta_{4}}} & {\color{rest}0} & {\color{second} y_{1}\frac{\delta}{\Delta_{4}}} & {\color{rest}0} & \cdots \\
\vdots & \vdots & \vdots & \vdots & \vdots & \vdots & \vdots & \ddots
\end{array} \right.\\[8mm]
& \qquad \qquad \qquad  \qquad \left. \begin{array}{cccccccc}
\ddots & \vdots & \vdots & \vdots & \vdots & \vdots & \vdots & \vdots \\
\cdots & {\color{rest}0} & {\color{first}\cJ_{n+m-5}} & {\color{rest}0} & {\color{first}0} & {\color{rest}0} & {\color{first}0} & {\color{first}0} \\
\cdots & {\color{second} -\cJ_{n+m-5}} & {\color{rest}0} & {\color{second}0} & {\color{rest}0} & {\color{second}0} & {\color{rest}0} & {\color{rest}0} \\
\cdots & {\color{rest}0} & {\color{first}y_{1}\Delta_{m-1}} & {\color{rest}0} & {\color{first}\cJ_{n+m-3}} & {\color{rest}0}& {\color{first}0} & {\color{first}0} \\
\cdots & {\color{second}y_{1}\frac{\delta}{\Delta_{m-1}}} & {\color{rest}0} & {\color{second} -\cJ_{n+m-3}} & {\color{rest}0} & {\color{second}0} & {\color{rest}0} & {\color{rest}0} \\
\cdots & {\color{rest}0} & {\color{first}\chi_{(n+m-3)}\Delta_{m-1}\Delta_{m}} & {\color{rest}0} & {\color{first}y_{1}\Delta_{m}} & {\color{rest}0} & {\color{first}\cJ_{n+m-1}} & {\color{first}0} \\
\cdots & {\color{second} -\chi_{(n+m-3)}\frac{\delta^{2}}{\Delta_{m-1}\Delta_{m}}} & {\color{rest}0} & {\color{second} y_{1}\frac{\delta}{\Delta_{m}}} & {\color{rest}0} & {\color{second} -\cJ_{n+m-1}}& {\color{rest}0} & {\color{rest}0} \\
\cdots & {\color{rest}0} & {\color{first}0} & {\color{rest}0} & {\color{first}\chi_{(n+m-1)}\Delta_{m}} & {\color{rest}0} & {\color{first}y_{1}} & {\color{first}\cJ_{n+m+1}}
\end{array}\right)
\end{align*}}}
\end{sideways}
\label{table:decomposition}
\caption{The upper right block of the matrix factorisation
$D_{(1)}\left(Q_{n(m)}\right)$ after the similarity transformations
$\cU^{f}_{n(m)}\dotsb \cU^{a}_{n(m)}$ -- it can be decomposed into two
parts: one (denoted in black) contains all entries in even lines (and
in the first one) and in odd
columns (and in the last one), and the other one (denoted in blue) consists of the
complement.}
\end{table}\afterpage{\clearpage}

\section{Deriving the alternative standard form}\label{sec:alternativeform}

In this section we want to sketch how we arrived at the alternative
standard form $\tilde{Q}_{n(m)}$ given
in~\eqref{eq:alternativestandardform}. Recall that we want to
successively apply $D_{(1)}$ on cones of polynomial factorisations with one
elementary factor $\cJ_{p_{i}}$ and decompose the result using similarity
transformations that leave the morphisms unchanged.

It turns out that to arrive at the alternative standard form, it is
enough to look at cones of three polynomial factorisations,
\begin{equation}
\mathcal{J}_{p,q,r}:=\begin{pmatrix}
\mathcal{J}_p & 0 & 0\\
1 & \mathcal{J}_{q} & 0\\
0 & 1 & \mathcal{J}_{r}
\end{pmatrix}\qquad (p,q,r\ \text{pairwise different})\ .
\end{equation}
If we now apply $D_{(1)}$ successively, the morphism entries $1$ will
be mapped to identity matrices. We then perform similarity
transformations and make sure that at the end the identity matrices
are untouched.

In the first step we find
\begin{equation}
D_{(1)}\big(  \mathcal{J}_{p,q,r}\big)=
\begin{pmatrix}
D_{(1)}\big( \mathcal{J}_p\big) & 0 & 0\\
\mathbb{1} & D_{(1)}\big( \mathcal{J}_{q}\big) & 0\\
0 & \mathbb{1} & D_{(1)}\big( \mathcal{J}_{r}\big)
\end{pmatrix}\ .
\end{equation}
We then blockwise transform $D_{(1)}\big( \mathcal{J}_n\big)$ to
$\tilde{\tilde{D}}_{0,m}\big( \mathcal{J}_n\big)$ as
in~\eqref{eq:deftildetildeD} to obtain
\begin{equation}\label{app:eq:D1onJpqrStd}
D_{(1)}\big( \mathcal{J}_{p,q,r}\big)\cong
\begin{pmatrix}
\tilde{\tilde{D}}_{0,1}\big( \mathcal{J}_p\big) & 0 & 0\\
\mathbb{1} & \tilde{\tilde{D}}_{0,1}\big( \mathcal{J}_{q}\big) & 0\\
0 & \mathbb{1} & \tilde{\tilde{D}}_{0,1}\big( \mathcal{J}_{r}\big)
\end{pmatrix}\ .
\end{equation}
When we now apply $D_{(1)}$ again, we already should have an idea,
which form we want to obtain for each block. As ingredients we take the
blocks $\tilde{D}_{0,m}\big(\cJ_{n}\big)$ that have the nice property
that their off-diagonal elements do not depend on $n$
(see~\eqref{eq:D0monJn}). 

In the standard form $\cJ_{n(m)}$ introduced in
section~\ref{sec:zeroell} we have on the diagonal the factors
$\cJ_{n'}$ with $n'$ going monotonically from $n-m$ to $n+m$. We now have
to reorder these entries, such that we can rewrite the expression in terms of the blocks
$\tilde{D}_{0,m'}\big(\cJ_{n}\big)$, which can be written as a cone of
$\cJ_{n+m'}$ and $\cJ_{n-m'}$ in either direction,
\begin{equation}
\begin{aligned}
\tilde{D}_{0,m}\big( \mathcal{J}_n\big)& =\cU_{m;n}^{(0)^{-1}}\cdot \begin{pmatrix}
\mathcal{J}_{n-m} & 0\\
y_1 & \mathcal{J}_{n+m}
\end{pmatrix}\cdot \cU^{(1)}_{m;n}\\
&=\cU_{m;n}^{(0)^{\dag^{-1}}}\cdot \begin{pmatrix}
\mathcal{J}_{n+m} & 0\\
y_1 & \mathcal{J}_{n-m}
\end{pmatrix}\cdot \cU_{m;n}^{(1)^{\dag}}
\end{aligned}
\end{equation}
where
\begin{equation}\label{app:eq:Umn}
\begin{split}
\cU_{m;n}&:=\cU_{\times}\left(2,2;\frac{1}{2\minus_{2m}\plus_{2m}},2\minus_{2m}\plus_{2m}\right)\cdot
\cU_{\times}\left(1,2;\frac{\plus_{2n-2m+1}}{\plus_{2n+2m+1}},\frac{\plus_{-2m-2n-1}}{\plus_{2m-2n-1}}\right)\cdot\\
&\qquad \cdot \cU_{col}\left(1,2;-\frac{y_1 \minus_{2n+2m+1}}{\plus_{2n+2m+1}}\right)\cdot
\cU_{row}\left(2,1;\frac{y_{1}\minus_{2n-2m+1}}{\plus_{2n-2m+1}}\right)\\
\cU_{m;n}^{\dag}&:=
\cU_{\times}\left(2,2;\frac{1}{2\minus_{2m}\plus_{2m}},2\minus_{2m}\plus_{2m}\right)\cdot
\cU_{\times}\left(1,2;
\frac{\plus_{-2m-2n-1}}{\plus_{2m-2n-1}},
\frac{\plus_{2n-2m+1}}{\plus_{2n+2m+1}}\right)\cdot\\
&\qquad \cdot 
\cU_{col}\left(1,2;\frac{y_{1}\minus_{2n-2m+1}}{\plus_{2n-2m+1}}\right)\cdot
\cU_{row}\left(2,1;-\frac{y_1 \minus_{2n+2m+1}}{\plus_{2n+2m+1}}\right)\\
\cU_{\times}(r,c;\alpha,\beta)&:=\cU_{col\times}(c;\beta)\cdot \cU_{row\times}(r;\alpha)\ .
\end{split}
\end{equation}
In particular, these transformations allow us to ``swap'' the
positions of any adjacent polynomial factors in our general formula for
$\cJ_{n(m)}$. Focusing only on the diagonal and first lower
sub-diagonal entries of $\cJ_{n(m)}$ for a moment, we can thus generate
from $\cJ_{n(m)}$ factorisations with the (sub-)diagonal entry
structure (the other lower diagonals have non-trivial entries)
\begin{equation}
\cJ_{n(m)}=\begin{pmatrix}
\mathcal{J}_{n-m} &  &  & & &\\
y_1 & \mathcal{J}_{n-m+2} &  & & &\\
& y_1 & \mathcal{J}_{n-m+4} &  && \\
 &  &\ddots  & &&\\
&& & &\mathcal{J}_{n+m-2} &  \\
&&&& y_1 &\mathcal{J}_{n+m}
\end{pmatrix}
\end{equation}
an alternative form, in which the (sub-)diagonal entries read (again the other lower diagonals have non-trivial entries)
\begin{equation}
\check{\cJ}_{n(m)}=\left\lbrace
\begin{array}{ccl}
\begin{pmatrix}
\mathcal{J}_{n} &  & && &\\
y_1 & \mathcal{J}_{n-2} &  && &\\
& y_1 & \mathcal{J}_{n+2} &  && \\
& & \ddots & &&\\
&& & &\mathcal{J}_{n-m} & \\
&&&& y_1 &\mathcal{J}_{n+m}
\end{pmatrix} & \qquad & \text{for $m\in 2\bZ$}\\
\begin{pmatrix}
\mathcal{J}_{n-1} & &&&&&\\
y_1 & \mathcal{J}_{n+1} &&&&&\\
& y_1 & \mathcal{J}_{n-3} &&&&\\
&        & y_1 & \mathcal{J}_{n+3} &&&&\\
&&& \ddots & &&\\
&&&& &\mathcal{J}_{n-m} & \\
&&&&& y_1 &\mathcal{J}_{n+m}
\end{pmatrix} & \qquad & \text{for $m\in 2\bZ+1$}
\end{array}
\right.
\end{equation}
and which may be obtained via suitable combinations of the
aforementioned similarity transformations. Unfortunately, in each
intermediate step, after an application of two similarity
transformations of the types listed in~\eqref{app:eq:Umn}, we need to
apply additional multiplicative transformations in order to ensure
that the entries on the sub-diagonal all read $y_1$ (i.e.\ with
constant prefactor $1$). Postponing the resolution of this
computational problem for the moment, we observe that once we have
transformed $Q_{n(m)}$ into the form $\check{Q}_{n(m)}$, we can
formulate yet another set of transformations (namely suitable inverse
transformations of type~\eqref{app:eq:Umn}) to express all diagonal blocks
in the form $\tilde{D}_{0,p}\big( \cJ_n\big)$ to obtain (note that
again the lower non-diagonal blocks are non-trivial)
\begin{equation}
\overline{\cJ}_{n(m)}:=\left\lbrace
\begin{array}{ccl}
\begin{pmatrix}
\mathcal{J}_{n} &  & & \\
  & \tilde{D}_{0,2}\big(  \mathcal{J}_{n}\big) &  & \\
& & \ddots & \\
&& & \tilde{D}_{0,m}\big(  \mathcal{J}_n\big)
\end{pmatrix} & \qquad & \text{for $m\in 2\bZ$}\\
\begin{pmatrix}
\tilde{D}_{0,1}\big( \mathcal{J}_{n}\big) & &&\\
 & \tilde{D}_{0,3}\big( \mathcal{J}_{n} \big) &&\\
& & \ddots & \\
&&& \tilde{D}_{0,m}\big(  \mathcal{J}_{n}\big)
\end{pmatrix} & \qquad & \text{for $m\in 2\bZ+1$.}
\end{array}
\right.
\end{equation}
Having described the general strategy, we can now go into the concrete computations.
The first step consists of computing
$D_{(1)}\big( D_{(1)}\big( \mathcal{J}_{p,q,r}\big)\big)$ explicitly, that is
via applying $D_{(1)}$ to $D_{(1)}\big(\mathcal{J}_{p,q,r}\big)$ in the
form~\eqref{app:eq:D1onJpqrStd}. Omitting the details of the
rather tedious computation (the computation can be done in the
framework of concatenations of fusion functors and will be presented
in a more general setting in~\cite{NBSFtwo2014}), we obtain the
decomposition
\begin{equation}
D_{(1)}\big( D_{(1)}\big( Q_{p,q,r}\big)\big) \cong Q_{p,q,r} \oplus
Q_{p,q,r(2)} \ ,
\end{equation}
where the summand $Q_{p,q,r(2)}$ is given by  
\begin{equation}
\mathcal{J}_{p,q,r(2)} = 
\begin{pmatrix}
\eta^4 \mathcal{J}_p  &                        0                       & 0 & 0 & 0 & 0\\
\Psi_{0,2}   & \tilde{\tilde{D}}_{0,2}\big( \mathcal{J}_p\big) &  0 & 0 & 0 & 0\\
1 & 0 & \eta^4 \mathcal{J}_q  &                        0            &   0 & 0  \\
0 & \mathbb{1}_{2\times 2} & \Psi_{0,2}   & \tilde{\tilde{D}}_{0,2}\big( \mathcal{J}_q\big) & 0 & 0\\
0 & 0 & 1 & 0 & \eta^4 \mathcal{J}_r  &                        0\\
0 & 0 & 0 & \mathbb{1}_{2\times 2} & \Psi_{0,2}   &
\tilde{\tilde{D}}_{0,2}\big( \mathcal{J}_r\big) 
\end{pmatrix}\ ,\qquad
\Psi_{0,2} =\begin{pmatrix}
y_1\\ \frac{1}{4\plus_{2}^{2}\plus_{4}}
\end{pmatrix}\ .
\end{equation}
We have thus achieved our goal of finding a new standard form
$\tilde{\cJ}_{n(2)}$ for the three diagonal blocks,
\begin{equation}
\tilde{\cJ}_{n(2)} = \begin{pmatrix}
\eta^{4}\cJ_{n} & 0\\
\Psi_{0,2} & \tilde{\tilde{D}}_{0,2}\big(\cJ_{n} \big)
\end{pmatrix} \ .
\end{equation}
In a similar fashion, using the similarity transformations of type~\eqref{app:eq:Umn}, we can determine the explicit formulae for
$\mathcal{J}_{p,q,r(3)}$ and $\mathcal{J}_{p,q,r(4)}$ by means of a long computation,\footnote{The main complication which makes these
computations difficult in practice is not so much the part of the
transformations necessary to transform each diagonal subblock of type
$\mathcal{J}_{n(m)}$, but rather to find those transformation
necessary in addition to bring the relative morphisms into the simple
form of $(m+1)\times (m+1)$ unit matrices. In particular, one encounters the
proliferation of rather complicated combinations of elementary
constants in the relative morphism entries.} with
the result that we find new standard forms $\tilde{\cJ}_{n(m)}$ for
the diagonal subblocks such that the morphisms between these
subblocks are simply unit matrices of size $(m+1)\times (m+1)$. 
The results we find for $m=3$ and $m=4$ fit into the following
inductive structure:
\begin{equation}
\begin{aligned}
m=1:\qquad \tilde{\cJ}_{n(1)} &= \tilde{\tilde{D}}_{0,1}\big( \mathcal{J}_n\big)\\
m=2:\qquad \tilde{\cJ}_{n(2)} &= \begin{pmatrix}
\eta^4 \mathcal{J}_n & 0\\
\Psi_{0,2} & \tilde{\tilde{D}}_{0,2}\big( \mathcal{J}_n\big)
\end{pmatrix}\,,&\; \Psi_{0,2}&=\begin{pmatrix}
y_1\\ \frac{1}{4\plus_{2}^{2}\plus_{4}}
\end{pmatrix}&\quad\\
m=3:\qquad \tilde{\cJ}_{n(3)} &= \begin{pmatrix}
\eta^4 \tilde{\cJ}_{n(1)} & 0\\
\Psi_{1,3} & \tilde{\tilde{D}}_{0,3}\big( \mathcal{J}_n\big)
\end{pmatrix}\,,&\; \Psi_{1,3}&=\begin{pmatrix}
\frac{1}{4\plus_{4}\plus_{-2}\plus_{2}} & y_1\\
0 & \frac{1}{4\plus_{4}\plus_{2}\plus_{6}}
\end{pmatrix}&\quad\\
m=4:\qquad \tilde{\cJ}_{n(4)} &= \begin{pmatrix}
\eta^4 \tilde{\cJ}_{n(2)} & 0\\
0_{2\times1}\Psi_{2,4} & \tilde{\tilde{D}}_{0,4}\big( \mathcal{J}_n\big)
\end{pmatrix}\,,&\; \Psi_{2,4}&=\begin{pmatrix}
\frac{1}{4\plus_{6}\plus_{-2}\plus_{4}} & y_1\\
0 & \frac{1}{4\plus_{6}\plus_{2}\plus_{8}}
\end{pmatrix}&\quad\\
\vdots\\
m=p:\qquad  \tilde{\cJ}_{n(p)} &= \begin{pmatrix}
\eta^4 \tilde{\cJ}_{n(p-2)} & 0\\
0_{2\times (p-3)} \ \Psi_{p-2,p} & \tilde{\tilde{D}}_{0,p}\big( \mathcal{J}_n\big)
\end{pmatrix}\,,\\
&\qquad \Psi_{p-2,p}=\begin{pmatrix}
\frac{1}{4\plus_{2p-2}\plus_{-2}\plus_{2p-4}} & y_1\\
0 & \frac{1}{4\plus_{2p-2}\plus_{2}\plus_{2p}}
\end{pmatrix}\ ,\mspace{-80mu}
\end{aligned}
\end{equation}
where for $m=4$ and in the last expression for $\tilde{\cJ}_{n(p)}$ we spelled out the size of the
zero-block in the lower left for clarity.

We conjecture that this structure holds for all $m$ up to a possible
truncation due to the finiteness of the level $k$. As we will discuss
in the following appendix~\ref{sec:finitelevelconstraints} we expect
the formula to be valid for $m+|n|\leq k/2$ if $k$ is even, and for
$m\leq k+2$ if $k$ is odd.

\section{Constraints at finite level}\label{sec:finitelevelconstraints}

We have to pay attention that all similarity transformations that we
perform are well-defined and that we do not accidentally divide by
zero. The coefficients we use,
$\plus_{p}$, $\minus_{p}$, $\chi_{(p)}$, $\Delta_{p}$ and $\gamma_{p,q}$
are generically neither zero or infinite, but there might be special
values where they lead to divergent expressions in the similarity
transformations.

The coefficients $\plus_{p}$ and $\minus_{p}$ (defined
in~\eqref{defplusminus}) are always finite, but
they can be zero:
\begin{equation}
\plus_{(2z+1)(k+3)} = 0 \quad (z\in\mathbb{Z})\quad ,\quad
\minus_{2z(k+3)} = 0\quad (z\in\mathbb{Z}) \ .
\end{equation}
The coefficients $\chi_{(p)}$ defined in~\eqref{defchi} can vanish,
\begin{equation}
\chi_{(p)}=0 \ \  \text{for}\ \ p=\left(\frac{k+2}{2}\pm 2  \right)
+z (k+3) \quad (z\in \mathbb{Z}) \ ,
\end{equation} 
and they can also diverge,
\begin{equation}
\chi_{(p)}=\infty \ \  \text{for}\ \ p=\left(\frac{k+2}{2}\pm 1 \right) 
+z(k+3) \quad (z\in \mathbb{Z}) \ .
\end{equation}
Note that $\chi_{(p)}$ is always regular and non-zero for odd level~$k$.

The coefficients $\Delta_{p}$ defined in~\eqref{app:defofDelta} can
also vanish,
\begin{equation}
\Delta_{p} = 0 \ \  \text{for}\ \ p=k+2 + z (k+3) \quad (z\in \mathbb{Z}) \ ,
\end{equation}
or diverge,
\begin{equation}
\Delta_{p} = \infty \ \  \text{for}\ \ p= z (k+3)  \quad (z\in \mathbb{Z}) \ .
\end{equation}
The analysis for $\gamma_{p,q}$ is a bit more complicated, but one can
show that it is regular and non-vanishing as long as
\begin{equation}
p,q\geq 0 \quad \text{and}\quad p+q \leq k+1 \ .
\end{equation}
We are now in the position to analyse when the similarity
transformations $\cU^{a},\dotsc ,\cU^{f}$ used in appendix~\ref{app:B}
to decompose
\begin{equation}
D_{(1)} \big(Q_{n(m)} \big) \cong Q_{n(m-1)}\oplus Q_{n(m+1)}
\end{equation}
are well-defined.

The transformations $\cU^{e}$ and $\cU^{f}$ are independent of the
label $n$ (see~\eqref{app:defUe} and~\eqref{app:defUf}). They contain
the inverse of $\Delta_{r}$ for $r=1,\dotsc ,m$ and also the inverse
of $\gamma_{r+1,m-r-1}$ for $r=1,\dotsc ,m-1$. From the considerations
above one finds that these quantities are well defined for $m\leq
k+1$.

The transformations $\cU^{a},\dotsc ,\cU^{d}$ contain inverses of
$\plus_{2p+1}$ and of $\chi_{(p)}$. One can observe immediately that
they can never be singular for odd level~$k$. For even~$k$, however,
we have to analyse the situation more carefully.
As an example look at the transformation $\cU^{d}$ (defined
in~\eqref{app:defUd}). It contains inverses of $\chi_{(p)}$ for
$p=n+m-2l-1$ where $l=0,\dotsc ,m-1$. For $m\geq 0$ the label $p$ satisfies
$|p|\leq m+|n|-1$. We have seen before that $\chi_{(p)}$ is regular
and finite for $|p|\leq \frac{k}{2}-2$, therefore all similarity
transformations are certainly regular for $m+|n|+1\leq
\frac{k}{2}$. One can show that this condition suffices to
guarantee that also the other transformations $\cU^{a}$, $\cU^{b}$ and
$\cU^{c}$ are regular.

We conclude that the formula for $Q_{n(m+1)}$ that we obtained from
the decomposition is valid if
\begin{equation}
\begin{array}{rll}
m+1+|n| &\leq \frac{k}{2} & \ \text{for}\ k\ \text{even}\\[1mm]
m+1 &\leq k+2 & \ \text{for}\ k\ \text{odd.}
\end{array}
\end{equation}
Similarly we can ask what the restrictions are on the alternative
standard form $\tilde{\cJ}_{n(m)}$ that is used for the factorisation corresponding to a
general boundary state $|L,\ell \rangle$. Because we use amongst
others the transformation to the the first standard form $\cJ_{n(m)}$, we
expect the constraint $m+|n|\leq k/2$ for even $k$ and $m\leq k+2$ for
odd $k$. One can check that also the additional transformations like
the transformations~\eqref{app:eq:Umn} to swap the entries on the
diagonal are well-defined if these conditions are satisfied. As the factorisation
for $|L,\ell\rangle$ is built from cones of
$\tilde{\cJ}_{0(\ell)},\dotsc,\tilde{\cJ}_{L(\ell)}$ we expect no
constraints for $k$ odd (because $\ell\leq k+1$ for all boundary
conditions), but for even $k$ we get the constraint $L+\ell\leq k/2$.
On the other hand we have the suspicion that one can also arrive at
the alternative standard form $\tilde{\cJ}_{n(m)}$ by blockwise similarity transformations
that do not depend on the label $n$. If this is true, then the
constraint could also not depend on the label $n$, and for even $k$ we would simply
obtain the constraint $m\leq k/2$. In that case the formula for the factorisation for
$|L,\ell\rangle$ would be correct for all $\ell\leq k/2$.

\section{A closed $2\times2$ form for $\cE_{n(m)}$ and $\cJ_{n(m)}$}\label{sec:closed}

In the main text we only considered the upper right block $Q^{(1)}$
(that we often denote by $\cJ$) of the matrix factorisations. The other
block $Q^{(0)}$ (that we often denote by $\cE$) can be reconstructed
from $Q^{(1)}$ by
\begin{equation}
Q^{(0)} = W_{2;k} \cdot \big( Q^{(1)}\big)^{-1} \ .
\end{equation}
Since the matrix factorisations $Q^{(1)}_{n(m)}$ have a simple triangular
structure, it is a straightforward recursive problem to determine the
inverse that we describe in the following. 

We start by writing explicitly the matrix elements of $\cJ_{n(m)}$,
\begin{equation}
\bJ_i{}^j\equiv \left(\cJ_{n(m)}\right)_i{}^j=\delta_{ij}\cJ_{n-m+2(i-1)}+\delta_{i-1,j}y_{1}+\delta_{i-2,j}\chi_{(n-m+2j)}\ ,
\end{equation}
where we introduced the shorthand notation $\bJ_i{}^j$ for the
components of $\cJ_{n(m)}$ for notational brevity in the ensuing
computations. The indices $i,j$ run from $1$ to $m+1$.

Introducing the additional shorthand notation
\begin{equation}
\bE_i{}^j\equiv \left(\cE_{n(m)}\right)_i{}^j\ ,
\end{equation}
we thus obtain an equation from which we can recursively determine the
structure of $\cE_{n(m)}$ (we write here and in the following $W\equiv W^{y}_{2;k}$ for brevity):
\begin{equation}
\begin{split}
\cJ_{n(m)}\cdot\cE_{n(m)}&=W\mathbb{1}\\
\Leftrightarrow\quad
\bJ_i{}^k\bE_k{}^j&=\left(\delta_{ik}\,\cJ_{n-m+2(k-1)}+\delta_{i-1,k}\,y_1+\delta_{i-2,k}\,\chi_{(n-m+2k)}\right)\bE_k{}^j
=W\delta_{ij}\ .
\end{split}
\end{equation}
Obviously, $\cE_{n(m)}$ is of lower triangular form,
\begin{equation}
\bE_{i}{}^{i+p}=0 \quad \text{for}\ p>0\ .
\end{equation}
The first non-trivial set of equations (for $i=j$) is
\begin{equation}
\begin{split}
&\cJ_{n-m+2(i-1)}\bE_i{}^i+\delta_{i-1,k}\,y_1\bE_k{}^i+\delta_{i-2,k}\,\chi_{(n-m+2k)}\bE_k{}^i
=W\\
\Leftrightarrow\quad &\bE_i{}^i=\frac{W}{\cJ_{n-m+2(i-1)}}\ ,
\end{split}
\end{equation}
where we used in the second line that $\delta_{i-1,k}\,\bE_k{}^i=\delta_{i-2,k}\,\bE_k{}^i=0$. The next special case is $i=j+1$, for which we obtain:
\begin{equation}
\begin{split}
&\left(\delta_{j+1,k}\,\cJ_{n-m+2(k-1)}+\delta_{j,k}\,y_1+\delta_{j-1,k}\,\chi_{(n-m+2k)}\right)\bE_k{}^j
=0\\\
\Leftrightarrow\quad &\bE_{j+1}{}^j=-\frac{y_1W}{\cJ_{n-m+2(j-1)}\cJ_{n-m+2j}}\ ,
\end{split}
\end{equation}
where we made use of the result $\delta_{j-1,k}\,\bE_k{}^j=0$ yet again.

\begin{figure}[htbp]
\begin{center}
\includegraphics[width=0.7\textwidth]{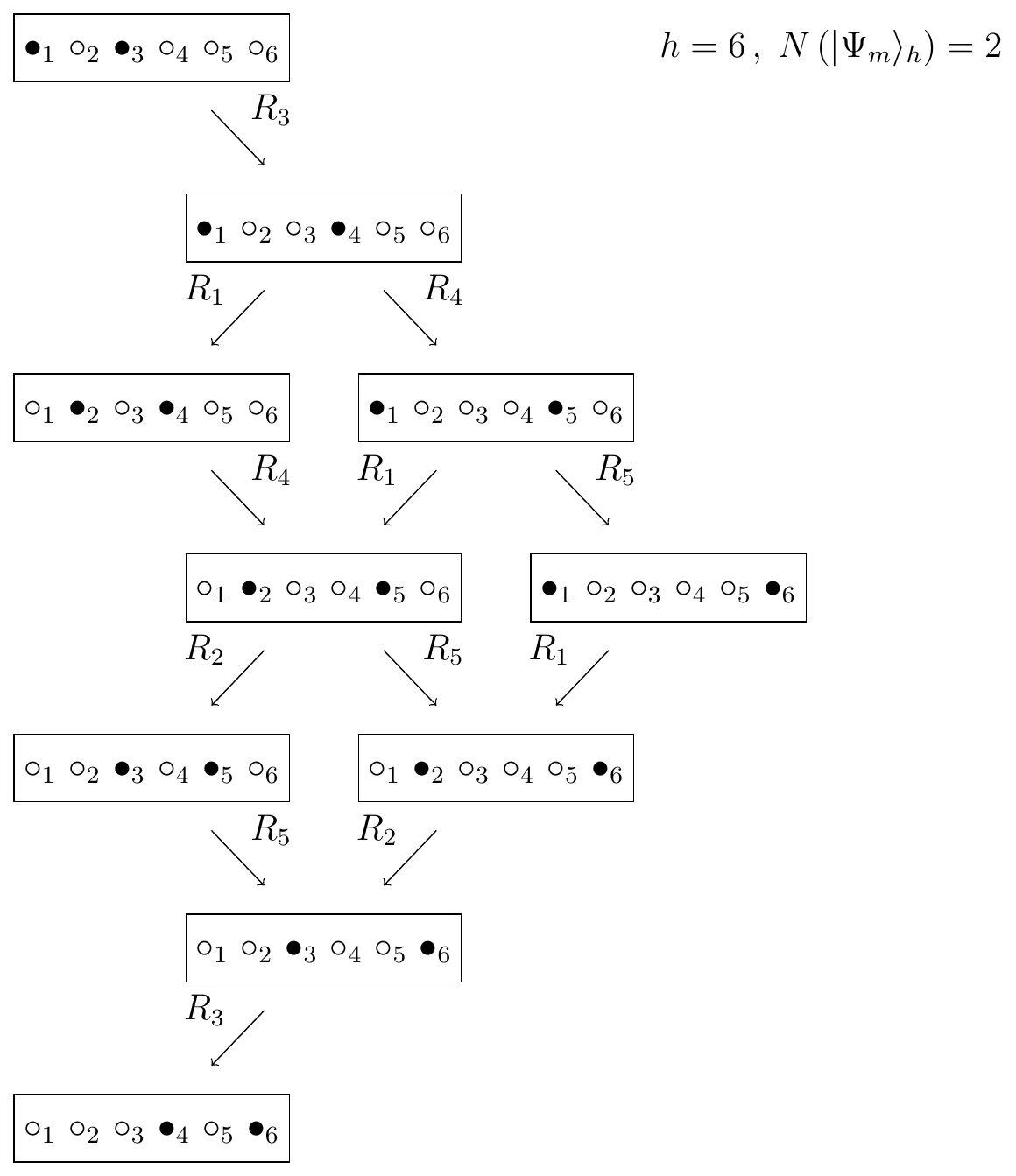}
\caption{The case $h=6$ and $N=2$.}
\label{fig:h6N2}
\end{center}
\end{figure}

\begin{figure}[htbp]
\begin{center}
\includegraphics[width=0.9\textwidth]{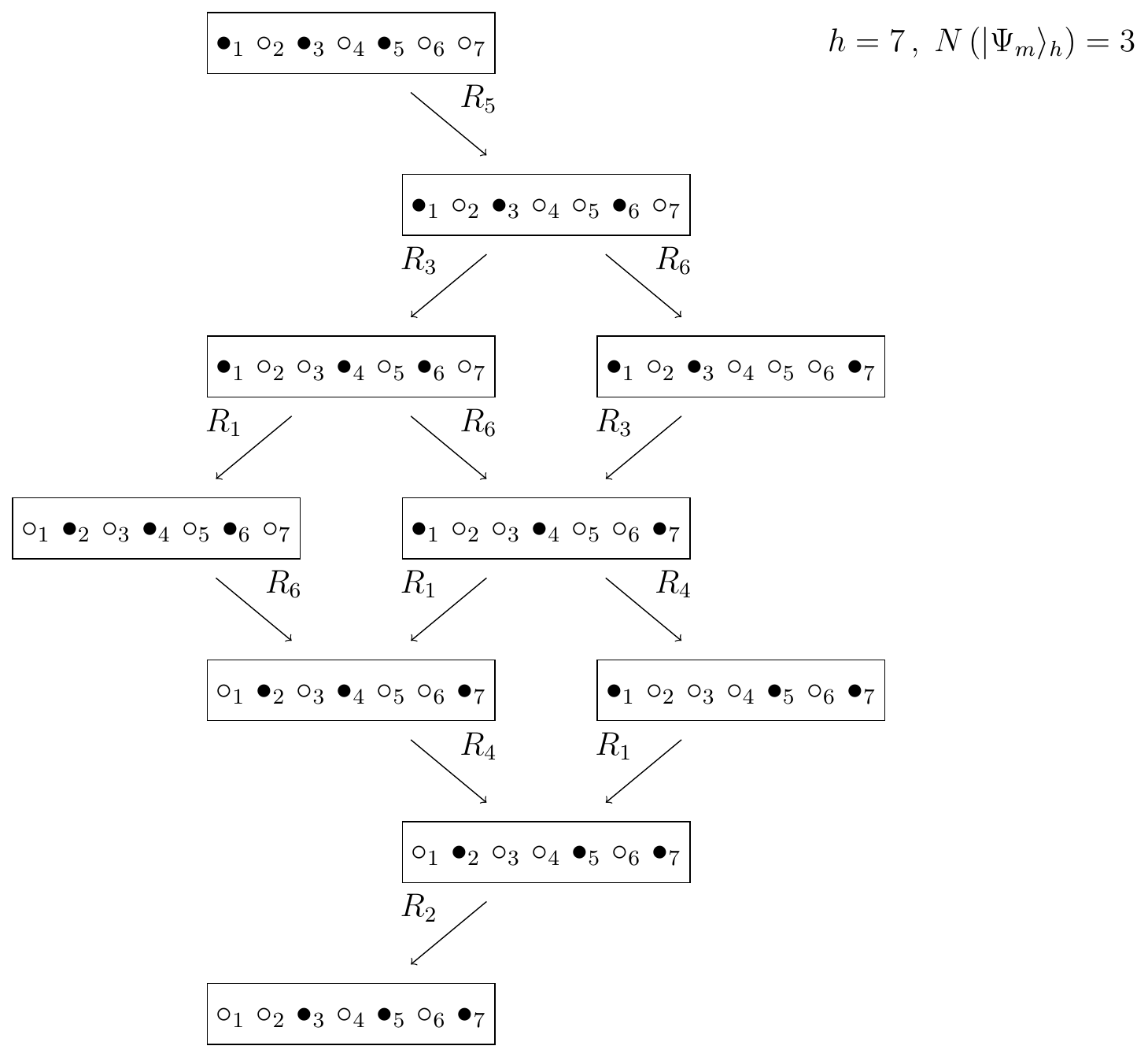}
\caption{The case $h=7$ and $N=3$.}
\label{fig:h7N3}
\end{center}
\end{figure}

For $i=j+2+p$ (with $p\in\mathbb{Z}_{\geq0}$), we obtain a double recursion relation
\begin{equation}
\begin{split}
&\left(\delta_{(j+2+p),k}\,\cJ_{n-m+2(j+p+1)}+\delta_{j+p+1,k}\,y_1+\delta_{j+p,k}\,\chi_{(n-m+2(j+p))}\right)\bE_k{}^j=0\\
\Leftrightarrow\quad &\boxed{\bE_{j+p+2}{}^j=-\frac{1}{\cJ_{n-m+2(j+p+1)}}\left(
y_1 \bE_{j+p+1}{}^j+\chi_{(n-m+2(j+p))}\bE_{j+p}{}^j
\right)} \ ,
\end{split}
\end{equation}
which relates the entries in the $p+2^{\text{nd}}$ lower diagonal to
the entries in the two diagonals above. Besides the dependence on the
recursion parameter $p$, the factors in the recursion relation only
depend on the combination $n-m+2j$, so we introduce 
the notation $\Psi_{l(p)}$ for the entries $\bE_{i}{}^{j}$ of
$\cE_{n(m)}$ defined by
\begin{equation}
\Psi_{n-m+2j+p-2\,(p+2)} := \big(\cE_{n(m)}\big)_{j+p+2}{}^{j}\ .
\end{equation}
The recursion relation then reads
\begin{equation}\label{app:recursion}
\Psi_{l(p)} = -\frac{1}{\cJ_{l+p}} \left(y_{1}\Psi_{l-1(p-1)} + 
\chi_{(l+p-2)}\Psi_{l-2(p-2)} \right) \ ,
\end{equation}
with 
\begin{equation}
\Psi_{l(1)} = -y_{1}\frac{W}{\cJ_{l-1}\cJ_{l+1}} \quad ,\quad 
\Psi_{l(0)} = \cE_{l} := \frac{W}{\cJ_{l}} \ .
\end{equation}
The first few solutions for $\cE_{n(m)}$ are given by
\begin{align}
\begin{split}
\cE_{n(0)}&=\bE_1{}^1\vert_{m=0,n=n}= \cE_{n}=\frac{W}{\cJ_n}
\end{split}\\
\begin{split}
\cE_{n(1)}&=\begin{pmatrix}
\cE_{n-1} & 0\\
\Psi_{n(1)} & \cE_{n+1}
\end{pmatrix}\;\Rightarrow\; \Psi_{n(1)}= \bE_{2}{}^1\vert_{m=1,n=n}=-\frac{y_1 W}{\cJ_{n-1}\cJ_{n+1}}
\end{split}\\
\begin{split}
\cE_{n(2)}&=\begin{pmatrix}
\cE_{n-2} & 0 & 0\\
\Psi_{n-1(1)} & \cE_{n} & 0\\
\Psi_{n(2)} & \Psi_{n+1(1)} & \cE_{n+2}
\end{pmatrix}\\
&\quad\Rightarrow\; \Psi_{n(2)}= \bE_{3}{}^1\vert_{m=2,n=n}=-\frac{1}{\cJ_{n+2}}\left(
y_1 \bE_{2}{}^1+\chi_{(n)}\bE_{1}{}^1
\right)\vert_{m=2,n=n}\\
&\quad\hphantom{\; \Psi_{n(2)}}=
-\frac{1}{\cJ_{n+2}}\left(y_1\left(-\frac{y_1 W}{\cJ_{n-2}\cJ_n}\right)+\chi_{(n)}\frac{W}{\cJ_{n-2}}\right)\\
&\quad\hphantom{\;
\Psi_{n(2)}}=-\frac{1}{\cJ_{n+2}}\left(y_1\Psi_{n-1(1)}+\chi_{(n)}\cE_{n-2}\right)\\
&\quad\hphantom{\; \Psi_{n(2)}}=
\frac{W}{\cJ_{n-2}\cJ_n \cJ_{n+2}}\left(y_1^2-\chi_{(n)}\cJ_n\right)
\end{split}\\
\begin{split}
\cE_{n(3)}&=\begin{pmatrix}
\cE_{n-3} & 0 & 0& 0\\
\Psi_{n-2(1)} & \cE_{n-1} & 0 & 0\\
\Psi_{n-1(2)} & \Psi_{n(1)} & \cE_{n+1} & 0\\
\Psi_{n(3)} & \Psi_{n+1(2)} & \Psi_{n+2(1)} &\cE_{n+3}
\end{pmatrix}\\
&\quad\Rightarrow\; \Psi_{n(3)}= 
\bE_{4}{}^1\vert_{m=3,n=n}=
-\frac{1}{\cJ_{n+3}}\left(
y_1 \bE_{3}{}^1+\chi_{(n+1)}\bE_{2}{}^1
\right)\vert_{m=3,n=n}\\
&\quad\hphantom{\; \Psi_{n(2)}}=-\frac{1}{\cJ_{n+3}}\left(y_1\Psi_{n-1(2)}+\chi_{(n+1)}\Psi_{n-2(1)}\right)\\
&\quad\hphantom{\; \Psi_{n(2)}}=
-\frac{y_1W}{\cJ_{n-3}\cJ_{n-1}\cJ_{n+1} \cJ_{n+3}}\left(y_1^2-\chi_{(n-1)}\cJ_{n-1}-\chi_{(n+1)}\cJ_{n+1}\right)
\end{split}\\
\vdots\notag
\end{align}
We want to obtain the general solution to the recursion
relation~\eqref{app:recursion}. We observe that in each recursion step
for $\Psi_{l(p)}$ we either go one step down in $p$ and pick up a
factor $-y_{1}/\cJ_{l+p}$ or we go two steps down in $p$ and pick up a
factor
\begin{equation}
-\frac{1}{\cJ_{l+p}}\chi_{(l+p-2)} = -\frac{1}{\cJ_{l+p}\cJ_{l+p-2}}
\chi_{(l+p-2)}\cJ_{l+p-2} \ .
\end{equation}
The recursion ends when we reach $p=0$. Therefore we can have at most
$\lfloor p/2\rfloor$ factors of $\chi$ in $\Psi_{l(p)}$. We call
$\Psi_{l(p)}^{(r)}$ the contribution to $\Psi_{l(p)}$ with $r$ factors of $\chi$,
such that
\begin{equation}
\Psi_{l(p)} = \sum_{r=0}^{\lfloor p/2\rfloor} \Psi_{l(p)}^{(r)} \ .
\end{equation}
When we go down always along the one-step recursion we have no factor
of $\chi$ and we get the contribution 
\begin{equation}
\Psi_{l(p)}^{(0)} = \prod_{j=0}^{p-1}\left(-\frac{y_{1}}{\cJ_{l+p-2j}} \right)
\frac{W}{\cJ_{l-p}} = W \prod_{j=0}^{p}\frac{1}{\cJ_{l+p-2j}} \left(-y_{1} \right)^{p}
 \ .
\end{equation}
When at one point we perform a two-step jump we get two factors of
$y_{1}$ less and instead a factor of
$-\chi_{(l-p+2m_{1})}\cJ_{l-p+2m_{1}}$ with $m_{1}=1,\dotsc ,p-1$
depending on where we do the two-step jump, so the contribution
$\Psi_{l(p)}^{(1)}$ is
\begin{equation}
\Psi_{l(p)}^{(1)} =  W \prod_{j=0}^{p}\frac{1}{\cJ_{l+p-2j}} \left(-y_{1} \right)^{p-2} 
\sum_{m_{1}=1}^{p-1} (-\chi_{(l-p+2m_{1})}\cJ_{l-p+2m_{1}}) \ .
\end{equation}
When we follow a two-step jump twice, we get again two factors of
$y_{1}$ less, and instead a factor of
$\chi_{(l-p+2m_{2})}\cJ_{l-p+2m_{2}}$ more. Note, however that the
difference of $m_{1}$ and $m_{2}$ has to be at least $2$ because of
course in a two-step jump we went \emph{two} steps down. These
arguments can easily be generalised to arbitrary numbers $r$ of
factors of $\chi$ and we find in total the result
\begin{equation}\label{app:solforPsi}
\Psi_{l(p)} = W \prod_{j=0}^{p}\frac{1}{\cJ_{l+p-2j}} 
\sum_{r=0}^{\lfloor p/2 \rfloor} \left(-y_{1} \right)^{p-2r} (-1)^{r}
\sum_{\substack{m_{1},\dotsc ,m_{r}\\ m_{i}+1<m_{i+1}\\ 1\leq m_{i}\leq p-1}} 
\prod_{i=1}^{r}\chi_{(l-p+2m_{i})}\cJ_{l-p+2m_{i}}  \ .
\end{equation} 
As an aside we mention a graphical way of organising the different
contributions. Introduce the analogue of a vacuum state for a spin chain, i.e.\ a state $|0\rangle_h$ with $h$ ``holes'', represented graphically as
\begin{equation}
|0\rangle_{h} \quad \widehat{=} \quad \boxed{\begin{matrix} \circ_1 & \circ_2 & \cdots & \circ_{h-1} & \circ_h\end{matrix}}\ .
\end{equation}
Then define the ``creation'' and ``annihilation'' operators $g^{+}_i$ and $g^{-}_i$ via
\begin{align}
g^{+}_i\boxed{\begin{matrix}
\cdots & *_i & \cdots
\end{matrix}}&:={\left\{\begin{array}{lcl}
\boxed{\begin{matrix}
\cdots & \bullet_i & \cdots
\end{matrix}} & \quad & \text{if $*_i=\circ_i$}\\
0 & \quad & \text{else}
\end{array}\right.}\\
g^{-}_i\boxed{\begin{matrix}
\cdots & *_i & \cdots
\end{matrix}}&:={\left\{\begin{array}{lcl}
\boxed{\begin{matrix}
\cdots & \circ_i & \cdots
\end{matrix}} & \quad & \text{if $*_i=\bullet_i$}\\
0 & \quad & \text{else.}
\end{array}\right.}
\end{align}%
Also, we need to implement the rule that we may never have two neighbouring ``excitations'' $\bullet_p\bullet_{p+1}$. Together with the preceding definitions, we may compactly express these requirements as ($\forall i$)
\begin{gather}
\begin{aligned}
g^{+}_i\circ g^{+}_i&=g^{-}_i\circ g^{-}_i=0\\
g^{+}_i\circ g^{-}_i&=g^{-}_i\circ g^{+}_i=id\\
g^{+}_i\circ g^{-}_{j}&=g^{-}_j\circ g^{+}_{i}\qquad (i\neq j)\\
g^{+}_i\circ g^{+}_{i+1}&=g^{+}_{i+1}\circ g^{+}_i=0\\
g^{+}_i\circ g^{+}_{i+2+p}&=g^{+}_{i+2+p}\circ g^{+}_i\qquad (p\geq 0)\ .
\end{aligned}
\end{gather}%
We may now define the operator $G$ as
\begin{equation}
G=G_1+G_2\,,\quad G_1:=\sum_{i=1}^{\left\lfloor \frac{h+1}{2}\right\rfloor}g^{+}_{2i-1}\,,\quad
G_2:=\sum_{i=1}^{h-1}R_i\ ,
\end{equation}
where 
\begin{equation}
R_i:=g^{-}_i\circ g^{+}_{i+1}
\end{equation}
is the operator that moves an ``excitation'' $\bullet_i$ one position to the right (unless of course if we have another ``excitation'' sitting at position $i+2$, in which case $R_i$ annihilates the given state). This allows us finally to generate all possible states $|\Psi\rangle_h$ via repeated action of the operator $G$ on the ``vacuum'' state $|0\rangle_h$ -- to this end, take the sum over arbitrary numbers of applications of $G$ (i.e.\ over $G^n$) applied to $|0\rangle_{h}$, and discard the multiplicities in this sum\footnote{It is obvious that only finitely many non-zero states can arise in this sum, since we only have finitely many sites in a given state $|0\rangle_h$, and thus we obtain only finitely many possibilities to excite a given vacuum state.}, 
\begin{equation}
P_h:=\left(\sum_{n=0}^{\infty} G^n|0\rangle_h\right)\bigg\vert_{\text{discard multiplicities}}=\sum_m |\Psi_m\rangle_h\ ,
\end{equation}
where $|\Psi_m\rangle_h$ denotes the inequivalent ``excited'' states. We will also need the operator $N$ which measures the number of ``excitations'' in a given state,
\begin{equation}
N(|\Psi_m\rangle_h)\equiv N\left(g^{+}_{m_1}\circ g^{+}_{m_2}\circ \dotsc\circ g^{+}_{m_p}|0\rangle_h\right):=\vert\{m_1,m_2,\dotsc\}\vert=p\ ,
\end{equation}
with $\vert\dotsc\vert$ denoting the cardinality of the set $\{m_1,m_2,\dotsc\}$. Finally, we define the \emph{evaluation operator}
\begin{equation}
\begin{split}
ev_n\left(|0\rangle_h\right)&:=1\\
ev_n\left(|\Psi_m\rangle_h\right)&\equiv ev_n\left(g^{+}_{m_1}\circ g^{+}_{m_2}\circ \dotsc\circ g^{+}_{m_p}|0\rangle_h\right)\\
&:=
\prod_{j=1}^{p}\chi_{(n-h-1+2m_j)}\cJ_{n-h-1+2m_j}\quad (p>0)\ .
\end{split}
\end{equation}
We can then rewrite the solution to $\Psi_{l(p)}$ as
\begin{equation}
\Psi_{l(p)} = W \prod_{j=0}^{p}\frac{1}{\cJ_{l+p-2j}} 
\sum_{r=0}^{\lfloor p/2 \rfloor} \left(-y_{1} \right)^{p-2r}
(-1)^{r}\,
 {}^{\,r}_{l}\!M_{p-1} 
\end{equation} 
with
\begin{equation}
{}^{\,r}_n{M}_h=\sum_{N(|\Psi_m\rangle_{h})=r}ev_n\left(|\Psi_m\rangle_h\right)\ .
\end{equation}
For example, the graphical representation of the set of states with
two ``excitations'' (i.e.\ $N=2$) at $h=6$ (together with the various
possibilities to generate the set $\{|\Psi_m\rangle_h\mid
N\left(|\Psi_m\rangle_h\right)=2\}$ from one of its representatives is
depicted in figure~\ref{fig:h6N2}, while figure~\ref{fig:h7N3}
represents the case $h=7$ and $N=3$.\\

The results above allow us actually to derive a compact
$2\times 2$ realisation of $\cE_{n(m)}$ and $\cJ_{n(m)}$. 
Consider the structure for $\cJ_{n(m)}$ as presented
in~\eqref{eq:rMF}. We may obviously choose to apply a number of row and column operations on $\cJ_{n(m)}$ in such a way that all rows and columns that intersect at a constant entry $\chi_{(p)}$ are ``cleared out'', to leave ultimately a form for $\cJ_{n(m)}$ of the form
\begin{align}
\cJ_{n(m)}&\cong\begin{pmatrix}
0 & 0 & \cdots &&&&&0&\widehat{\cJ}_{n(m)_{11}}&\widehat{\cJ}_{n(m)_{12}}\\
0 &0 & 0 & \cdots &&&&0&\widehat{\cJ}_{n(m)_{21}}&\widehat{\cJ}_{n(m)_{22}}\\
\chi_{(n-m+2)} & 0& 0& 0 & \cdots &&&0&0&0\\
0 & \chi_{(n-m+4)} & 0 &0& 0 & \cdots &&\vdots&\vdots&\vdots\\
\vdots & \ddots & \ddots & \ddots & \ddots & \ddots & &&&\\
\\
&&&&&& 0 & \chi_{(n+m-2)} & 0 & 0
\end{pmatrix}\label{eq:rMFv2}\\
&\cong \left(\cJ_{triv}\right)^{\oplus (m-1)}\oplus \widehat{\cJ}_{n(m)}\ .
\end{align}%

\noindent Now, the structure of $\widehat{\cJ}_{n(m)}$ will generically become very complicated for large values of $m$, but we claim that
\begin{equation}\label{app:JEW}
\widehat{\cJ}_{n(m)} \widehat{\cE}_{n(m)} = W\cdot \one_{2\times 2} 
\end{equation}
where $\widehat{\cE}_{n(m)}$ is the lower left $2\times 2$ subblock of $\cE_{n(m)}$,
\begin{equation}\label{app:hatE}
\widehat{\cE}_{n(m)}:=\begin{pmatrix}
\Psi_{n-1(m-1)} & \Psi_{n(m-2)}\\
\Psi_{n(m)} & \Psi_{n+1(m-1)}
\end{pmatrix}\ .
\end{equation}
We will prove this statement below. This result also allows us to give
an explicit result for the $2\times 2$ matrix $\widehat{\cJ}_{n(m)}$
as 
\begin{equation}
\widehat{\cJ}_{n(m)}=W\left(\widehat{\cE}_{n(m)}\right)^{-1}=\frac{W}{\det
\widehat{\cE}_{n(m)}}\begin{pmatrix}
\Psi_{n+1(m-1)} & -\Psi_{n(m-2)}\\
-\Psi_{n(m)} & \Psi_{n-1(m-1)}
\end{pmatrix}\ .
\end{equation}
The determinant of $\widehat{\cE}_{n(m)}$ can be obtained as
follows. From the form of $\cJ_{n(m)}$ in~\eqref{eq:rMF} it is obvious
that
\begin{equation}
\det \cJ_{n(m)} = \prod_{j=0}^{m}\cJ_{n-m+2j} \ .
\end{equation}
When we perform the column and row manipulations to obtain the form~\eqref{eq:rMFv2}, we do not change the determinant\footnote{Note that to arrive at~\eqref{eq:rMFv2} we only performed transformations where we added multiples of rows (colums) to other rows (columns) and we did not rescale any row (column), so that the determinant remains unchanged.}, so
from~\eqref{eq:rMFv2} we see that
\begin{equation}
\det \cJ_{n(m)} = \det \widehat{\cJ}_{n(m)} \cdot
\prod_{j=1}^{m-1}\chi_{n-m+2j} \ .
\end{equation} 
On the other hand, according to~\eqref{app:JEW}
\begin{equation}
\det \widehat{\cJ}_{n(m)} \cdot \det \widehat{\cE}_{n(m)} = W^{2} \,,
\end{equation} 
which leads to
\begin{equation}
\frac{W}{\det \widehat{\cE}_{n(m)}} = \frac{1}{W}
\frac{\prod_{j=0}^{m}\cJ_{n-m+2j}}{\prod_{j=1}^{m-1}\chi_{(n-m+2j)}}\ .
\end{equation}
Our final result for $\widehat{\cJ}_{n(m)}$ is then
\begin{equation}
\widehat{\cJ}_{n(m)}=\frac{1}{W}
\frac{\prod_{j=0}^{m}\cJ_{n-m+2j}}{\prod_{j=1}^{m-1}\chi_{(n-m+2j)}}
\begin{pmatrix}
\Psi_{n+1(m-1)} & -\Psi_{n(m-2)}\\
-\Psi_{n(m)} & \Psi_{n-1(m-1)}
\end{pmatrix}\ .
\end{equation}

It remains to prove the claim~\eqref{app:JEW}. To this end, consider
the induced effect of a given row or column transformation of the
$\cJ$-block on the $\cE$-block of a
matrix factorisation $Q$. According to~\eqref{eq:Ulist}, performing
e.g.\ a similarity transformation which adds row $r$ times a
polynomial $p$ to row $s$ of the $\cJ$-block leads to a transformation
of the $\cE$-block in which the column $s$ multiplied by $-p$ is added
to the column $r$ of the $\cE$-block. If we now choose to start
constructing the transformation from the form $\cJ_{n(m)}$ to the form
$\widehat{\cJ}_{n(m)}$ by ``clearing out'' the entries above the
constant entries $\chi_{(p)}$ with a number of row operations, we
first of all observe that the subblock of $\cE_{n(m)}$ that
corresponds to $\widehat{\cE}_{n(m)}$ remains unaltered. Similarly,
afterwards performing a number of column operations on the
$\cJ_{n(m)}$ block, inducing row transformations on the $\cE_{n(m)}$
block, will not affect the $\widehat{\cE}_{n(m)}$ subblock. In
summary, what we have obtained so far is that one may find a set of
similarity transformations that brings the $\cJ_{n(m)}$ block into the
form $\widehat{\cJ}_{n(m)}$ without affecting the
$\widehat{\cE}_{n(m)}$ subblock of the $\cE_{n(m)}$ block. Now, due to
the fact that
\begin{equation}
\cJ_{n(m)}\cong \widehat{\cJ}_{n(m)}\oplus \cJ_{triv}^{\oplus m-1}\,,
\end{equation}
we automatically must have
\begin{equation}
\cE_{n(m)}\cong \widehat{\cE}_{n(m)}\oplus \cE_{triv}^{\oplus m-1}\,,
\end{equation}
which concludes the proof.


\bibliographystyle{mystyle5}
\bibliography{refs}

\end{document}